# ALGORITHMIC OPACITY AS A PRINCIPAL-AGENT PROBLEM

IGNACIO COFONE* & KATHERINE J. STRANDBURG**

*Principal-agent problems are central to debates over algorithmic transparency, but have been underexplored. Concerns about gaming and trade secrecy, frequently invoked to justify opacity, can also mask self-serving strategic behavior by decision-makers, who may prioritize private benefits over social welfare. This article develops a principal-agent test for determining when algorithmic decision-making should trigger disclosure obligations even when concerns about gaming or trade secrecy exist. It proposes that disclosure requirements should be anchored in decision quality.*

*The article introduces structured patterns of error types and the distribution of decision outcomes across groups as diagnostic tools for detecting principal-agent misalignment—extending their role beyond fairness metrics. These make visible whether decision-makers align algorithmic design and implementation with the public interest or instead use opacity to obscure bias, inefficiency, or strategic shirking.*

* Professor of Law and Regulation of AI, University of Oxford, Faculty of Law and Institute for Ethics in AI. ignacio.cofone@law.ox.ac.uk. We thank Nikita Aggarwal, Ifeoma Ajunwa, Jack Balkin, Claudia Haupt, Marcel Kahan, Helen Nissenbaum, Ben Green, Dan Solove, Orla Lynskey, Thomas Nachbar, Shu-Yi Oei, Paul Ohm, Leanna Pennington, Jennifer Pinsof, Alicia Solow-Niederman, Salome Viljoen, Ari Waldman, Rebecca Williams, and Tal Zarsky for their comments on earlier drafts. This article also benefited from comments received at the Privacy Law Scholars Conference, the Imperfect Enforcement Conference at Yale Law School, the Emory Global AI and Law Colloquium, and invited presentations at Boston College Law School, Seoul National University, Torcuato Di Tella University, NYU School of Law's Engelberg Center and Information Law Institute, the Yale Law School Information Society Project. We especially thank Nicholas Tilmes for his outstanding research assistance throughout the writing process and Anthony Perrins and Tobit Glenhaber for their excellent assistance with finalizing citations and sources.

** Pauline Newman Professor of Law and Faculty Director, Information Law Institute, NYU School of Law. katherine.strandburg@nyu.edu. Professor Strandburg acknowledges the generous support of the Filomen D'Agostino and Max E. Greenberg Research Fund.

*Strong disparities across social groups, and asymmetries in the distribution of decision errors across outcome types, should trigger rebuttable presumptions favoring further disclosure obligations.*

*This principal-agent framework reorients regulatory attention toward institutional design questions that transparency debates have not focused on so far. This framework informs the adoption of differentiated disclosure regimes: staged, partial, mediated, or audience-segmented. By reframing opacity debates through the lens of principal-agent theory, this article challenges long-standing assumptions about algorithmic secrecy: it shows the domain of justifiable algorithmic opacity is significantly narrower than commonly assumed once differentiated disclosure regimes are considered. When gaming or trade secrecy risks are legitimate, actionable disclosures (such as training data characteristics and key system design parameters) ordinarily can be mandated without compromising innovation or enabling gaming. This approach offers an implementable regulatory pathway for AI transparency in both public and private sectors.*

*By developing burden-shifting presumptions about what type of opacity is impermissible, what minimum disclosures are required, and how to weigh trade-secrecy and gaming claims, the article provides policymakers and courts with a framework for structuring tailored disclosure orders, evaluating secrecy claims, and allocating burden of proof in transparency disputes. The analysis also shows that meaningful disclosures are possible even for technically inscrutable algorithms.*

## INTRODUCTION

Algorithmic transparency has an elephant in the room. Even before "black box" artificial intelligence (AI) was widely debated, Amazon developed a machine learning CV-screening algorithm that systematically disadvantaged women.[1] Although the AI screening promised to enhance neutrality and nondiscrimination, it was highly biased.[2] This episode illustrates a recurring pattern: algorithmic systems frequently reproduce existing failures, so regulating AI systems is often an exercise in identifying new manifestations of old legal problems. Deliberate opacity (namely, the intentional lack of transparency) is one of them.

This instance did not just uncover a problem at Amazon; rather, it illustrates the broader concern that opaque algorithmic decision-making systems can mask errors, entrench bias, and obscure discrimination, all while simultaneously amplifying (or at least sustaining) them. A considerable body of literature discusses the benefits and limitations of transparency and disclosure for providing accountability for algorithmic decision-making systems.[3] Much of the existing discussion focuses on the technical opacity of some "black box" algorithms and their tendency to be inexplicable even to their designers.[4] But the importance of technical opacity is often overstated, as substantial information can be disclosed about even the most technically opaque machine learning algorithm.[5]

[1] Jeffrey Dastin, *Amazon Scraps Secret AI Recruiting Tool That Showed Bias Against Women*, REUTERS (Oct. 10, 2018), https://www.reuters.com/article/idUSKCN1MK0AG/ (on file with the NYU Journal of Intellectual Property and Entertainment Law).

[2] *Id.* (Amazon tried to automate hiring with a machine learning algorithm, but upon testing it realized that it merely perpetuated the tech industry's bias against women). *See also* SAFIYA UMOJA NOBLE, ALGORITHMS OF OPPRESSION: HOW SEARCH ENGINES REINFORCE RACISM 181 (2018) ("In essence, we need greater transparency and public pressure to slow down the automation of our worst impulses. We have automated human decision making and then disavowed our responsibility for it. Without public funding and adequate information policy that protects the rights to fair representation online, an escalation in the erosion of quality information to inform the public will continue.").

[3] *See generally, e.g.*, Joshua A. Kroll et al., *Accountable Algorithms*, 165 U. PA. L. REV. 633 (2017); ANDREW G. FERGUSON, THE RISE OF BIG DATA POLICING: SURVEILLANCE, RACE, AND THE FUTURE OF LAW ENFORCEMENT (2017); John Zerilli et al., *Transparency in Algorithmic and Human Decision-Making: Is There a Double Standard?*, 32 PHIL. & TECH. 661 (2019); Cary Coglianese & David Lehr, *Regulating by Robot: Administrative Decision Making in the Machine-Learning Era*, 105 GEO. L.J. 1147 (2017); Tal Z. Zarsky, *Transparent Predictions*, 2013 U. ILL. L. REV. 1503 (2013).

[4] *See generally, e.g.*, Andrew D. Selbst et al., *Deconstructing Design Decisions: Why Courts Must Interrogate Machine Learning and Other Technologies*, 85 OHIO ST. L.J. 415, 416–17 (2024) ("Black boxes are ubiquitous in narratives surrounding accountability for technological harms, especially machine learning and artificial intelligence. The black box is usually the villain of the story, the reason that there cannot be accountability for the harms. The black box is a trade secret. It's inscrutable. It's both."); Danielle Keats Citron & Frank Pasquale, *The Scored Society: Due Process for Automated Predictions*, 89 WASH. L. REV. 1

Instead, this article focuses on *deliberate* opacity. Deliberate opacity is routinely excused either by claims of trade secrecy or by asserted fears that decision-subjects (i.e., individuals whose lives are affected by a decision-making algorithm, such as loan applicants or job candidates) will "game the system."[6] Our earlier work pointed out the limits of decision-subjects' power to game even the most transparent algorithmic decision-making systems.[7] We presented a framework for recognizing the limited circumstances under which disclosure about a decision-making system is likely to facilitate decision-subject gaming.[8] But decision-subject gaming captures only part of the strategic landscape surrounding algorithmic transparency.

Throughout this Article, algorithmic decision-making system encompasses predictive AI (including machine learning models that classify or score decision-subjects), generative AI (including large language models that produce recommendations, assessments, or decisions as text), and agentic AI (including systems that autonomously execute decision sequences on behalf of an operator). The principal-agent framework developed here applies across all three types

(2014); Frank Pasquale, The Black Box Society: The Secret Algorithms That Control Money and Information (2015); Andrew D. Selbst & Solon Barocas, *The Intuitive Appeal of Explainable Machines*, 87 Fordham L. Rev. 1085 (2019); Noble, *supra* note 2; Virginia Eubanks, Automating Inequality: How High-Tech Tools Profile, Police, and Punish the Poor (2018).

[5] *See* Rishi Bommasani et al., *The Foundation Model Transparency Index*, arXiv (Oct. 19, 2023), https://arxiv.org/abs/2310.12941 [https://perma.cc/XFR6-MQNJ]. *See also infra* Section II.C.

[6] *See, e.g.*, Charles T. Graves & Sonia K. Katyal, *From Trade Secrecy to Seclusion*, 109 Geo. L.J. 1337, 1341 (2021) ("[A]ssertions of trade secrecy … extend to different types of information as well, as companies learn that labeling sensitive or embarrassing information as a 'trade secret' or 'confidential' can stall or silence calls for disclosure"); Rebecca Wexler, *Life, Liberty, and Trade Secrets: Intellectual Property in the Criminal Justice System*, 70 Stan. L. Rev. 1343, 1371–1377 (2018) (detailing justifications for trade secrecy in criminal law).

[7] Ignacio Cofone & Katherine J. Strandburg, *Strategic Games and Algorithmic Secrecy*, 64 McGill L.J. 623, 634 (2019).

[8] Here and in our earlier work, we define gaming as socially undesirable strategic behavior that rewards decision-subjects (people about whom algorithmic systems made decisions) with unwarranted positive outcomes. Our definition does not include socially desirable strategic behavior, such as using information about the algorithm to improve one's eligibility for a favorable decision or to strategically avoid an unjustified negative decision.

because the problem—decision-makers using opacity to avoid accountability for their choices—arises whenever an AI system mediates consequential decisions.

Interactions between decision-subjects and those involved in designing, developing, and implementing algorithmic decision-making systems—like interactions between decision-subjects and human decision-makers—are strategic on both sides.[9] None of these players should be assumed to be acting in society's best interest.

This Article addresses the strategic behaviors of decision-makers (including both developers and deployers of decision-making systems) with regard to opacity. We argue that in public and private contexts decision-makers are, to varying degrees, agents for the public. Principal-agent relationships are undermined when agents act strategically to promote their own interests at the expense of the principal's goals. The agents involved in algorithmic decision systems are no exception.[10] We show how claims about gaming risk and trade secrecy can shield strategic behavior that is contrary to public interest. We argue that managing these principal-agent problems should be a primary consideration in the design of effective disclosure regulation. This means that decision-makers' claims about gaming and trade secrecy should not be accepted at face value. Moreover, because disclosure is a crucial mechanism for detecting principal-agent problems,[11] its value may outweigh its costs even when concerns about gaming or trade secrecy have merit.

---

[9] Benjamin Laufer et al., *Strategic Evaluation: Subjects, Evaluators, and Society*, EQUITY & ACCESS ALGORITHMS, MECH., & OPTIMIZATION 1 (Oct. 30, 2023), https://doi.org/10.1145/3617694.3623237; Jack M. Balkin, *How to Regulate (and not Regulate) Social Media*, KNIGHT FIRST AMEND. INST. OCCASIONAL PAPERS, 13–15 (Mar. 25, 2020) [hereinafter Balkin, *How to Regulate (and not Regulate) Social Media*]; Zachary Lipton, *The Mythos of Model Interpretability*, 61 COMMUNICATIONS OF THE ACM, 36 https://doi.org/10.1145/3233231; Ignacio Cofone, *Servers and Waiters: What Matters in the Law of A.I.*, 21 STAN. TECH. L. REV. 167 (2018). *See also* Jack Balkin, *2016 Sidley Austin Distinguished Lecture on Big Data Law and Policy: The Three Laws of Robotics in the Age of Big Data*, 78 OHIO ST. L.J. 1217, 1226 (2017) [hereinafter Balkin, *The Three Laws of Robotics*].

[10] *See, e.g.*, Jon Kleinberg & Manish Raghavan, *How Do Classifiers Induce Agents to Invest Effort Strategically?*, 8 ACM TRANSACTIONS ON ECON. & COMPUTATION 19:1, 19:15 (2020).

[11] Noam Kolt, *Governing AI Agents*, 101 NOTRE DAME L. REV. (forthcoming 2026) (noting the role of disclosure duties in principal-agent relations, and arguing they are complicated by the difficulty to interpreting AI models' decisions).

When decision-subject gaming is likely or when trade secrecy claims are plausible, disclosing details about a decision-making algorithm involves tradeoffs.[12] On the one hand, disclosure provides accountability and allows the principal (here, the public) to supervise the agent's activities. However, disclosure can be socially costly for two distinct reasons. First, gaming is costly for decision-subjects in terms of time and effort that is wasted, rather than employed productively. It is costly when it makes the decision-making algorithm less informative because some decision-subjects receive undeserved benefits, requiring decision-makers to either engage in further screening efforts or make worse decisions. Second, exposing trade secrets may also have social costs in terms of innovation, though those costs must be balanced against the extent to which disclosure promotes follow-on innovation.

While opacity can inhibit decision-subject gaming and protect trade secrets, it can also mask socially undesirable strategic choices about algorithm design and implementation, while preventing socially valuable uses of information by decision-subjects and competitive improvers.[13] When decision-makers argue that secrecy is necessary to avoid undesirable gaming or prevent inappropriate competitor free riding, they may be acting strategically. This uncertainty counsels in favor of structured oversight mechanisms to supervise the designers, developers, procurers, and deployers of algorithmic decision systems.[14] Failure to account for these principal-agent problems deprives society not only of disclosure's immediate benefits to decision-subjects, but also of the improvements in decision quality that could result from disclosure that bolsters accountability.[15] So, decisions about what and when to disclose should not be left to decision-makers' discretion: the potential costs of disclosure must be balanced against its social, not merely private, benefits.

Disclosure is not an all-or-nothing choice. Disclosures about decision-making systems implicate multiple *audiences*, whose varying interests and potential uses of the information require different social balances.[16] One potential audience is *decision-subjects*, who can use certain information about a decision-making system strategically for both socially beneficial and socially deleterious purposes. Another is the public writ large, for whom disclosure provides information about decision quality that can be used to oversee decision-makers acting on their behalf and to ensure that incentives are aligned with the public interest. A third audience is composed of commercial competitors, who can use information about a decision-making system either to free ride on a previous innovator's investment or to develop socially valuable follow-on innovation. Because these audiences use information about a decision-making system for different purposes, they are interested in different (though possibly overlapping) types of information. And the audiences themselves overlap: decision-subjects and potential competitors are also members of the public. Because of this, conceiving of disclosure as an all-or-nothing choice limits one to socially sub-optimal approaches.

[12] *See infra* Section I.E.

[13] *See infra* Part III.

[14] *See* Ignacio Cofone, The Privacy Fallacy: Harm and Power in the Information Economy 59–66 (2023) (discussing the particular importance of accountability mechanisms in the information economy).

[15] *See, e.g.*, Ganesh Ghalme et al., *Strategic Classification in the Dark*, 139 Proc. 38th Int'l Conf. on Mach. Learning (2021) (finding transparency can bolster accuracy based on the gap in prediction error between opaque and transparent but strategy-robust proxies); Qiaochu Wang et al., *Algorithmic Transparency with Strategic Users*, 69 Mgmt. Sci. 2297 (2022) (finding that making ML algorithms transparent may increase their predictive power despite decision-subject gaming even when gaming is low effort).

This means that there are tradeoffs involved. The relevant tradeoffs vary by information type, audience, and context, so disclosure should be structured as a *differentiated disclosure regime*.[17] Disclosure mandates should attempt to account for the variety of information that could be disclosed, as well as the contextual nuances of the tradeoffs involved. Here, we outline a normative framework for designing a disclosure regime that accounts for strategic behavior on both sides and for the varying social interests associated with disclosing to different audiences.[18]

We show that error rates, systematic error patterns, and the distribution of decision outcomes across groups can function as evidentiary signals of principal-agent misalignment, indicating where principal-agent problems are likely to be distorting algorithm design, and the likely effects of disclosure on decision-subject strategic behavior and innovation.[19] Because this information can be disclosed without raising concerns about gaming or trade secrecy, it provides a low-cost trigger for further scrutiny. If these clues raise red flags, decision-makers can be required to explain and justify the observed error rate patterns and additional disclosure can be mandated through staged, burden-shifting mechanisms. While this use of error rates aligns with fairness metrics proposed in the literature, [20] our proposal treats them as diagnostic triggers for further disclosure and investigation, thereby avoiding many of the difficulties associated with using fairness metrics as evidence of illegal discrimination or embedding them into algorithmic systems.

[16] *See* Sue Lim & Ralf Schmalzle, *The Effect of Source Disclosure on Evaluation of AI-Generated Messages*, 2 COMPUT. HUM. BEHAV. 1, 8 (2024), https://doi.org/10.1016/j.chbah.2024.100058. *See also* Charles R. Korsmo, *The Audience for Corporate Disclosure*, 102 IOWA L. REV. 1581, 1632 (2017).

[17] *See infra* Part IV. *See also* Graves & Katyal, *supra* note 6, at 1412 ("Trade secret protection should not be viewed as a monolith where the skimpiest satisfaction of the elements of trade secrecy means that regulatory or other disclosure in the public interest is impossible. There are contexts where the public interest in disclosure is strong, and the case for competitive harm is weak. The challenge is to articulate simple and flexible tests courts can employ to protect trade secrets where harms are real but also to separate instances where disclosure is appropriate.").

[18] *See infra* Part IV.

The article proceeds as follows. Part I reviews and critiques the most often asserted rationales for keeping decision-making systems secret: gaming, trade secrecy, and technical opacity. It also briefly describes the social benefits of disclosing information about such algorithms. Parts II and III explain how and when these rationales for opacity can be deployed strategically by decision-makers to the detriment of society. Part II examines the reasons why we treat decision-makers in public and some private contexts as agents of the public and gives an overview of the sorts of principal-agent problems that might be expected to occur. Part III introduces the concept of an error profile (i.e., the distribution of error rates across groups) and analyzes how error profiles can provide important clues to whether principal-agent problems are likely to infect proxy design. Finally, Part IV develops implementable regulatory design, exploring how to mitigate these principal-agent problems by designing disclosure mechanisms that appropriately balance the social costs of disclosure with its social benefits.

## I
## Opacity Rationales

New York City's school admissions lottery assigns students to schools through a matching algorithm. The probability for an individual student to be matched with a particular school depends on a combination of priority rules and random numbers ("lottery numbers") in a process that is not disclosed to parents or students.[21] School officials justify this secrecy by a combination of trade secrecy considerations and broad concerns that, if information were disclosed, parents would game the system.[22]

[19] *See infra* Part III.

[20] *See generally* Binnur Uçar et al., *Algorithmic Fairness: A Tolerance Perspective*, arXiv (May 17, 2024), https://arxiv.org/abs/2405.09543 [https://perma.cc/Y5EM-DV7G] (analyzing fairness metrics according to "the degree to which variations in fairness outcomes are acceptable"); Alessandra Castelnovo et al., *A Clarification of the Nuances in the Fairness Metrics Landscape*, 12 Sci. Rep. 4209 (2022) (providing a taxonomy of fairness metrics).

The opacity of the school lottery process is typical. Decision-makers give three types of justifications for keeping algorithmic systems opaque. First, they justify intentional secrecy as necessary to avoid undesirable "gaming" by decision-subjects; in a prior article, we explored the limited circumstances in which this concern is plausible.[23] Second, they assert trade secrecy, particularly (but not exclusively) by or on behalf of commercial developers of decision-making systems. Third, they explain that the algorithms created by some machine learning techniques are often technically inscrutable, meaning that mathematical and computational representations of the algorithm are not completely understandable by humans, even if they have the relevant technical expertise.[24] These justifications emphasize the private and social costs of disclosure but often ignore its social benefits.

In this Part, we briefly discuss each of these rationales. We then canvass the social benefits of disclosure, which should be weighed against these costs.

### *A. The Gaming Trope*

Decision-makers often contend that disclosure of decision-making criteria is "undesirable, such as when it discloses private information or permits tax cheats or terrorists to game the systems determining audits or security screening."[25] This

---

[21] *See* Amelie Marian, *Algorithmic Transparency and Accountability through Crowdsourcing: A Study of the NYC School Admission Lottery*, 6 PROC. 2023 ACM CONF. ON FAIRNESS, ACCOUNTABILITY, & TRANSPARENCY 434, 435 (2023). *See also* Amelie Marian, *Results from the 2024 NYC School Admission Lottery Surveys*, MEDIUM (Mar. 15, 2024), https://medium.com/algorithms-in-the-wild/results-from-the-2024-nyc-school-admission-lottery-surveys-7b1a6910987c [https://perma.cc/AR5Q-FYQG].

[22] *See* Amelie Marian, *Algorithmic Transparency and Accountability through Crowdsourcing: A Study of the NYC School Admission Lottery*, 6 PROC. 2023 ACM CONF. ON FAIRNESS, ACCOUNTABILITY, & TRANSPARENCY 434, 436 (2023).

[23] Cofone & Strandburg, *supra* note 7.

[24] Selbst et al., *supra* note 4, at 417; Sylvia Lu, *Algorithmic Opacity, Private Accountability, and Corporate Social Disclosure in the Age of Artificial Intelligence*, 23 VAND. J. ENT. & TECH. L. 99, 115–16 (2020).

[25] Kroll et al., *supra* note 3, at 633–39 ("The process for deciding which tax returns to audit, or whom to pull aside for secondary security screening at the airport, may need to be partly opaque to prevent tax cheats or

argument assumes an audience of decision-subjects. Thus, one potential way to avoid gaming is to disclose gameable aspects of a decision-making system to some sort of auditor or oversight body. The specter of gaming is raised in a range of situations, yet the implied assumption that the expected social costs of gaming outweigh the benefits of disclosure is rarely elaborated.[26] This assumption depends on the difficulty and cost of gaming, as detailed in earlier work.[27]

Significant gaming is usually possible only if a decision-making system employs relatively weak proxies (i.e., stand-ins) for the criteria that its decision-maker would ideally like to measure. As explored in more depth below, highly faithful proxies are generally difficult to game because there is usually some underlying causal connection between the algorithmic output and the decision criteria it is intended to represent.[28] Even a noisy (low-quality) proxy can be relatively impervious to gaming. The choice of system objectives, training data, and model type (in predictive systems) or system architecture (more broadly), are entirely in the system designer's hands and thus impossible to game. A decision-subject also cannot game characteristics that are overly costly or impossible to change, such as age, prior criminal record, or religious affiliation.[29] Moreover, terrorists from gaming the system. When the decision being regulated is a commercial one, such as an offer of credit, transparency may be undesirable because it defeats the legitimate protection of consumer data, commercial proprietary information, or trade secrets. Finally, when an explanation of how a rule operates requires disclosing the data under analysis and those data are private or sensitive (e.g., in adjudicating a commercial offer of credit, a lender reviews detailed financial information about the applicant), disclosure of the data may be undesirable or even legally barred."); Jane Bambauer & Tal Zarsky, *The Algorithm Game*, 94 Notre Dame L. Rev. 1, 10 (2019) (formalizing the decision-subject gaming concern, though not delving into the question of when disclosure can and will lead to undesirable gaming).

[26] *See* Nicholas Diakopoulos, *Accountability, Transparency, and Algorithms*, *in* The Oxford Handbook of Ethics of AI, 197, 205–06 (Markus D. Druber et al. eds., 2020) (acknowledging fears of "gaming and manipulation, understandability, privacy, temporal instability, sociotechnical intermingling, costs, competitive concerns, and legal contexts" but arguing that they "should be understood less as undermining the premise of transparency than as moderators that must be taken into account in order to design and configure an effective implementation of algorithmic transparency for any specific context"); Kroll et al., *supra* note 3, at 658.

[27] Cofone & Strandburg, *supra* note 7, at 643–57. Still, well-designed classifiers can incentivize agents to invest effort in improving their outcomes rather than gaming. *See* Kleinberg & Raghavan, *supra* note 10, at 19:15.

[28] Cofone & Strandburg, *supra* note 7, at 636.

[29] *But see* Andrew Estornell et al., *Group-Fair Classification with Strategic Agents*, Proc. 2023 ACM Conf. on Fairness, Accountability, & Transparency 389 (2023) (finding that some decision subjects misrepresent such traits to take advantage of fair machine learning techniques).

and especially for systems with complex or opaque mappings between inputs and outputs (such as machine learning systems incorporating many variables), it will be unclear what combinations of changes in one's characteristics, if any, will improve eligibility for a favorable decision, and costly to coordinate the necessary suite of changes.[30]

As a result, decision-subjects will only game the system if there are individual characteristics they can strategically manipulate enough to change the outcome at a low enough private cost. Moreover, the situations in which such manipulations constitute gaming are even more limited. Gaming is strategic behavior that makes decision-subjects appear more eligible for beneficial decisions than they are. Decision-subjects cannot meaningfully be said to be "gaming the system" when they respond to disclosure strategically by changing their behavior in socially desirable ways, most relevantly by improving their true eligibility for a beneficial decision.[31] There are also situations in which a decision-subject's strategic response to disclosure can correct for an algorithm's biases or other mistakes, thus producing a socially preferable result.[32] Strategic error correction and eligibility improvement are socially beneficial consequences of disclosure that may offset the social costs of gaming.

In sum, a given disclosure cannot seriously increase the threat of socially undesirable gaming unless four prerequisites are met.[33] First, proxies for the ideal decision-making criteria (the socially preferred basis for the decision, whether grounded in law, normative priorities, or policy objectives) are sufficiently imperfect that there is enough "wiggle room" for gaming. Second, the proposed disclosure pertains to characteristics that are (sufficiently) modifiable by decision-subjects in ways that improve their chances of a beneficial decision. Third, modifying those characteristics is cost-effective for the decision-subject. Fourth, modifying those characteristics improves the proxy without improving or better reflecting the decision-subject's true eligibility for a beneficial decision. Thus, even if gaming is a socially relevant concern, there is usually a considerable amount of information about a decision-making system that can be disclosed without running the risk of decision-subject gaming.

---

[30] *See, e.g.*, Mareike Möhlmann et al., *Algorithmic Management of Work on Online Labor Platforms: When Matching Meets Control*, 45 MIS QUARTERLY 1999 (2021) (noting that after "[Uber] drivers used online communities as a vehicle to form coalitions and 'game' the system . . . Uber was able to identify some of the loopholes and constantly updated the system in order to make it more difficult for drivers to game the system to their advantage").

[31] *See* Flavia Barsotti et al., *Transparency, Detection and Imitation in Strategic Classification*, PROC. 31ST INT'L JOINT CONF. ON A.I. 67 (2022); Emilee Rader et al., *Explanations as Mechanisms for Supporting Algorithmic Transparency*, INT'L CONF. ON HUM. FACTORS COMPUT. SYS. 1 (Apr. 19, 2018).

[32] At the same time, robust de-biasing measures may reduce incentives for strategic manipulation. *See* Xueru Zhang et al., *Fairness Interventions as (Dis)Incentives for Strategic Manipulation*, PROC. 39TH INT'L CONF. ON MACHINE LEARNING (2022). *See also* Julia Angwin et al., *Machine Bias*, PROPUBLICA (May 23, 2016), https://www.propublica.org/article/machine-bias-risk-assessments-in-criminal-sentencing [https://perma.cc/DT4C-WDJE].

[33] Cofone & Strandburg, *supra* note 7, at 663.

### *B. The Trade Secrecy Trope*

Some decision-makers argue that disclosing information about a decision-making algorithm is undesirable because it will reveal valuable trade secret information.[34] Proponents of the trade secrecy objection argue that companies invest significant resources in developing proprietary algorithms and decision-making systems, and that exposing the details of those systems enables competitors to free-ride on those investments, diminishing their market share and profitability and consequently decreasing ex-ante incentives to invest in innovative decision proxy development.[35] These concerns are, for example, explicitly considered in Freedom of Information Act Requests, which allow agencies to withhold information that may interfere with trade secrecy.[36] While trade secrecy claims are most common in entirely commercial contexts, they are also made in the public arena when public entities procure an automated decision-making algorithm from private companies.[37] Further, some public entities even claim trade secrecy for their home-grown algorithms. Public agencies invoke trade-secrecy protections for similar reasons as private ones: although they do not compete in commercial markets, they rely on private vendors whose contracts incorporate confidentiality terms, and they occasionally assert secrecy over in-house tools to avoid oversight.

---

[34] Margot A. Kaminski, *Regulating the Risks of AI*, 103 B.U. L. Rev. 1347, 1408–09 (2023); Sonia K. Katyal, *Private Accountability in the Age of Artificial Intelligence*, 66 UCLA L. Rev. 54, 121–25 (2019); Jenna Burrell, *How the Machine "Thinks": Understanding Opacity in Machine Learning Algorithms*, 3 Big Data & Soc'y 1, 3–4 (2016). *See also* Neil M. Richards & Jonathan H. King, *Three Paradoxes of Big Data*, 66 Stan. L. Rev. Online 41, 43 (2013).

[35] *See, e.g.*, Bo Cowgill & Catherine Tucker, *Algorithmic Fairness and Economics*, Colum. Bus. Sch. Rsch. Paper (Feb. 14, 2020), https://dx.doi.org/10.2139/ssrn.3361280; Joshua New & Daniel Castro, *How Policymakers Can Foster Algorithmic Accountability*, Ctr. for Data Innovation (May 21, 2018), https://datainnovation.org/2018/05/how-policymakers-can-foster-algorithmic-accountability/ [https://perma.cc/7ZGF-JSP4]. *See also* Wexler, *supra* note 6, at 1421 (providing an overview of the argument).

[36] Bernard W. Bell et al., *Disclosure of Agency Legal Materials*, Rep. to Admin. Conf. U.S. (June 2, 2023); Coglianese & Lehr, *supra* note 3, at 1210; Katherine Fink, *Opening the Government's Black Boxes: Freedom of Information and Algorithmic Accountability*, 21 Info. Commc'n & Soc'y 1453 (2018).

[37] Cary Coglianese, *Procurement and Artificial Intelligence*, *in* Handbook on Public Policy and AI 235, 235–36 (Regine Paul, Jennifer Cobbe & Emma Carmel, eds., Edward Elgar, 2024); Paul B. de Laat,

Trade secrecy is not intended to be an all-purpose mechanism for stifling competition or rewarding market actors, however.[38] It has two primary social justifications. First, like other forms of intellectual property, it arguably incentivizes innovation by preventing competitors from free-riding on investments in developing new technology, thus providing innovators a period of market exclusivity to recoup those investments.[39] In addition, it is sometimes justified as a means for policing markets against misappropriation. As prior work details, [40] the misappropriation justification is irrelevant to mandatory disclosure, so in the present context, the persuasiveness of the trade secrecy justification for non-disclosure rides on its purported benefit of promoting innovation by deterring undesirable free-riding.

In general, market competition is socially valuable because it reduces prices and promotes innovation. Disclosure can facilitate these social benefits by reducing wasteful investments in duplicative efforts and facilitating follow-on innovation.[41] Trade secrecy is socially desirable only to the extent that disclosure would render innovators unable to recoup their free-rideable investments.[42] However,

---

*Algorithmic Decision-Making Employing Profiling: Will Trade Secrecy Protection Render the Right to Explanation Toothless?*, 24 ETHICS & INFO. TECH. 17, at 9 (2022).

[38] Wexler, *supra* note 6, at 1423 ("[A]s compared to substantive trade secret doctrine, the trade secret evidentiary privilege overprotects intellectual property and contradicts the disclosure-prompting purpose of trade secret law, and even the broader goal of encouraging innovation fails to justify a total withholding remedy under the privilege.").

[39] *See* Daniel J. Hemel & Lisa L. Ouellette, *Innovation Policy Pluralism*, 128 YALE L.J. 544, 567–68 (2019).

[40] Eli Siems et al., *Trade Secrecy and Innovation in Forensic Technology*, 73 HASTINGS L.J. 773, 796–98 (2022).

[41] *See, e.g.*, Travis A. Dyer et al., *The Effect of Patent Disclosure Quality on Innovation*, 77 J. ACCT. & ECON. 101647 (2024), https://doi.org/10.1016/j.jacceco.2023.101647; Jinhwan Kim & Kristen Valentine, *The Innovation Consequences of Mandatory Patent Disclosures*, 71 J. ACCT. & ECON. 101381 (2021), https://doi.org/10.1016/j.jacceco.2020.101381.

[42] Siems et al., *supra* note 40, at 796 ("Reverse engineering and independent invention . . . are viewed as crucial limitations on trade secrecy's downstream social costs."); Amy Kapczynski, *The Public History of Trade Secrets*, 55 U.C. DAVIS L. REV. 1367, 1396 (2022) ("The beneficial incentives to innovate created by trade secret law, for example, may not outweigh the costs of diverting creators from using the patent

innovators often can recoup those investments in ways that do not require secrecy. For example, many publicly developed decision-making systems are not part of a competitive market but are financed directly by in-house development or a government procurement process.[43] Even where competitive free-riding is a potential issue, first-mover advantages, network effects, and other forms of intellectual property, such as copyright or patent, may be sufficient for innovators to recoup their free-rideable investments.[44] Many technology acquisitions also function like prize systems in that the winner obtains a contractually (or practically) ensured extended period of exclusivity that is sufficient to incentivize innovation.[45]

The innovation justification for trade secrecy is often particularly weak in the context of decision proxies. To begin with, there is no justification for trade secrecy in completely non-commercial contexts, such as when government decision-makers design their proxies.[46] This is because, in that situation, the costs of innovation are paid by the public directly, rather than being privately paid and then recouped on the market. Even in many cases in which government decision-makers procure decision proxies from commercial entities, which is a common approach, the innovation rationale for secrecy may be weak.[47] If the procurement involves

---

system (which awards term-limited protection and ensures public disclosure), and of duplicative research (particularly likely where learning is kept secret).”). *See also* Robert G. Bone, *The (Still) Shaky Foundations of Trade Secret Law*, 92 Tex. L. Rev. 1803, 1809 (2014).

[43] *See generally* Merve Hickok, *Public Procurement of Artificial Intelligence Systems: New Risks and Future Proofing*, 39 AI & Soc’y 1213, (2024) (collecting examples of decision-making systems financed through government procurement); *see also* Coglianese, *supra* note 37, at 239–40.

[44] *See, e.g.*, *id.*; Charlotte A. Tschider, *Legal Opacity: Artificial Intelligence’s Sticky Wicket*, 106 Iowa L. Rev. Online 126 (2021). *See also* Coglianese, *supra* note 37, at 240.

[45] Lingling Suo, Kezhi Yang & Hao Ji, *The Impact of Technological Mergers and Acquisitions on Enterprise Innovation: A Review*, 15 Sustainability no. 17:12883 (2023), https://doi.org/10.3390/su151712883 (explaining that technology mergers and acquisitions are a strategy for acquiring and internalizing external technologies in order to strengthen the acquirer’s technological leadership and sustain its competitive advantage); Xinao Li, *The Impact of Technology Mergers and Acquisitions on Enterprise Sustainable Competitiveness*, 16 Sustainability no. 6:2291 (2024), https://doi.org/10.3390/su16062291 (finding that acquiring technological assets boosts the acquirer’s sustainable competitiveness).

[46] Graves & Katyal, *supra* note 6, at 1381 (arguing governmental entities’ assertion of trade secrecy “creates a foundational conflict between commercial interests in secrecy for competitive advantage and the tenets of transparency that we normally associate with government accountability. When government trade secrets are asserted, rather than being used as a sword in a misappropriation action … they ‘have always been used as a shield to public disclosure.’ ”).

[47] *Id.* at 1377–78; Ryan Calo & Danielle K. Citron, *The Automated Administrative State: A Crisis of Legitimacy*, 70 Emory L.J. 797, 823–26 (2021).

an ex-ante contract for a bespoke design, payment terms are negotiated directly to incorporate any expected costs of innovation.[48] In such procurement settings, the developer is already compensated for its innovation through the contract itself, so secrecy is not required to ensure recovery of research and development costs, as the incentive has been priced into the agreement. Justified concerns about disclosing trade secrets therefore arise only when a decision proxy is developed and deployed in a commercial context or when a public body commercially procures an "off the shelf" (or mostly off the shelf) proxy.

Even where there is a plausible argument that a commercial entity requires a mechanism to recoup innovative investments before free-riding competitors can swoop in to undercut its profits, unlimited trade secrecy is usually not socially desirable.[49] Despite popular descriptions of trade secrecy as providing "indefinite" protection, that is not ordinarily the case—nor would unlimited trade secrecy be socially optimal.[50] Trade secrecy protection, which was developed before the era of data-driven innovation, has two important limiting exceptions: independent invention and reverse engineering.[51] The expectation, which is sensible for most industrial innovations, is that the time it takes for competitors to reverse engineer or independently come up with an innovation is roughly calibrated to the investment required to produce it.

Decision-making systems often do match this standard expectation. They may be nearly impossible to reverse engineer because of the way they are deployed. Like standard industrial processes, they are used in secret, often by a single entity. Unlike standard industrial processes, however, which produce products that are sold on the market and can be examined and analyzed by competitors, decision-making systems produce sets of individual decisions, the results of which are spread over many decision-subjects and often can only be evaluated by the decision-making entity. This lack of information makes reverse engineering by competitors extremely difficult if not impossible.[52] When the proxy is data-driven, this situation is often exacerbated because a particular decision-maker has unique access to the data that was used to create the proxy. This control over relevant data not only makes reverse engineering even more difficult but also undercuts the possibility of independent invention.

[48] *See* Coglianese, *supra* note 37, at 2 ("[T]he secrecy surrounding the work of private contractors that develop and deploy these algorithms is not inherent to the technology .... Governments can require, through provisions in their contracts, that AI service providers release information to satisfy the public and courts that these tools are functioning fairly and responsibly.")

[49] *See* de Laat, *supra* note 37, at 16–17; Camilla A. Hrdy & Mark A. Lemley, *Abandoning Trade Secrets*, 73 Stan. L. Rev. 1, 60 (2021) (arguing unlimited trade secrecy might make us reasonably be more concerned about employers gaming the system by belatedly claiming plans to use the secret after the employee takes it).

[50] *See* Hrdy & Lemley, *supra* note 49, at 66 ("[T]he end of a trade secret's life ... encourages the dissemination of information that could not otherwise be known or developed. And it frees up space for employee-inventors, who can use secrets that their original owners have given up on. The public can benefit from abandoning trade secrets."). *See also* Camilla A. Hrdy, *Keeping ChatGPT a Trade Secret While Selling It Too*, 40 Berkeley Tech. L.J. 75 (2025).

[51] Unif. Trade Secrets Act § 1 cmt. at 6 (amended 1985) ("[I]f reverse engineering is lengthy and expensive, a person who discovers the trade secret through reverse engineering can have a trade secret in the information obtained from reverse engineering."). *See, e.g.*, Andrew A. Schwartz, *The Corporate Preference for Trade Secret*, 74 Ohio St. L.J. 623, 630 (2013).

Ordinarily, more difficult reverse engineering and independent invention is a justification for longer trade secrecy, under the assumption that such difficulty indicates that developing the technology required correspondingly high levels of investment.[53] That connection is often broken for decision-making proxies, however, because the private information needed is often relatively cheap for the decision-maker to acquire, but prohibitively expensive for competitors to obtain.[54] Information about what factors are important to a decision, as well as about how well a given version of the proxy is functioning, is often uniquely available to the decision-makers who deploy it. Those decision-makers can opt to share the information exclusively with a selected proxy developer (either within their commercial organization or through a procurement process). In some cases, either the ultimate decision-maker or the commercial developer also has exclusive access to relevant input datasets that were collected as a byproduct of other activity.[55] If much of the information needed to develop the proxy is already in the decision-maker's hands, the extra investment required to create and hone the proxy may be relatively low.

[52] In addition, some AI models' terms of use seek to prohibit attempts at reverse engineering based on a purported trade secrets rationale. *See* Hrdy, *supra* note 50, at 118 ("[A]ll ChatGPT users are subject to an anti-reverse-engineering clause that prohibits a wide variety of methods of reverse engineering ChatGPT's secrets.").

[53] *See* W. Nicholson Price II, *Expired Patents, Trade Secrets, and Stymied Competition*, 92 Notre Dame L. Rev. 1611, 1621 (2017) (noting that, compared to patents, "[t]he trade secret trade-off is different: no expiration, and potentially lower costs, but a narrower scope and the ability of competitors to invent-around or reverse-engineer"); Hrdy, *supra* note 50, at 154–55.

[54] *See* Herbert Hovenkamp, *Antitrust and Platform Monopoly*, 130 Yale L.J. 1952, 2033–35 (2021); D. Daniel Sokol & Roisin Comerford, *Antitrust and Regulating Big Data*, 23 Geo. Mason L. Rev. 1129, 1147–48 (2016).

[55] Sokol & Comerford, *supra* note 54, at 1133–35.

Innovation policy (and intellectual property doctrine) ordinarily attempts to balance the financial incentives required to induce innovation with the social value of allowing competitors access to information that will jumpstart follow-on innovation.[56] In patent law, for example, this balance is reflected in the oft-cited "quid pro quo" represented by the exchange of patent exclusivity for nearly immediate public access to enabling information about the invention. Because the doctrines that limit the scope of trade secrecy (which are often criticized by the literature in any event) are frequently dysfunctional for decision-making proxies, innovation considerations alone counsel skepticism about demands by commercial decision proxy developers for broad or unlimited trade secrecy.[57] Other approaches, such as staged disclosure and alternative mechanisms for recouping innovative investments, are likely to be socially preferable.[58]

Finally, market-based approaches to inducing innovation, such as trade secrecy and intellectual property generally, only work when purchasers can assess the quality of the products they are buying so that the profits recouped by the innovator reflect the social value of the innovation.[59] Otherwise, there is a market failure. There are many existing regulatory regimes, ranging from nutrition labeling to pre-market drug approvals to professional licensing regimes that aim to address this sort of market failure. This failure affects what economists call "credence goods," which are goods that are difficult or impossible for purchasers to value accurately even after using them.[60] Decision-making AI systems, in many ways, operate as credence goods because those who use them might not know their quality even while or after they apply them.

---

[56] Hemel & Ouellette, *supra* note 39, at 544 (describing intellectual property as "a combination of two distinct elements: an innovation incentive that promises a market-based reward to producers of knowledge goods, and an allocation mechanism that makes access to knowledge goods conditional upon payment of a proprietary price").

[57] *See, e.g.*, Case C-203/22, CK v. Magistrat der Stadt Wien, ECLI:EU:C:2025:117, ¶ 67–76 (Feb. 2. 2025).

[58] Aaron Springer & Steve Whittaker, *Progressive Disclosure: When, Why, and How Do Users Want Algorithmic Transparency Information?*, 10 ACM TRANSACTIONS ON INTERACTIVE INTELLIGENT SYS. 29, 29:24–29:26 (2020). *See generally* Deepa Muralidhar, *The Effect of Progressive Disclosure in the Transparency of Explainable Artificial Intelligence Systems*, IEEE SYMP. ON VISUAL LANGUAGES & HUMAN-CENTRIC COMPUT. 382 (2024).

[59] *See* Jan Biermann et al., *Algorithmic Advice as a Credence Good*, ZEW DISCUSSION PAPERS NO. 22-071 (2022); Loukas Balafoutas & Rudolf Kerschbamer, *Credence Goods in the Literature: What the Past Fifteen Years Have Taught Us About Fraud, Incentives, and the Role of Institutions*, 26 J. BEHAV. & EXPERIMENTAL FIN. 100285 (2020).

This Article focuses on situations in which the ultimate "purchaser" of a decision proxy is the public. Without sufficient disclosure, it is difficult for the public to reliably assess the social value of a decision-making system. If the public cannot independently assess its social value and cannot rely on the decision-maker to act as a faithful agent, then even if trade secrecy incentivizes some sort of investment, it cannot be trusted to incentivize socially optimal innovation.[61] A disclosure regime that enables the public to assess decision quality is needed to align decision-makers' incentives with society's values and remedy the market failure.[62] In sum, a commercial market for decision proxy development can only function properly when there is sufficient disclosure to permit the public to oversee the decision-maker's performance as its agent.

In the commercial context, disclosure's role in aligning innovation with social value may be in some tension with trade secrecy's goal of allowing developers to recoup innovative investments before their profits are undercut by free-riding competitors. However, two points are worth noting. First, there is no point in incentivizing the wrong innovation: there is a public interest in decision quality, so disclosure regimes for decision proxies must be designed with this dual purpose in mind. Second, alternative mechanisms for recouping innovative investments that do not rely on secrecy should be employed when possible.[63]

In sum, while there may be social benefits to keeping some proprietary information relating to decision-making systems away from potential commercial competitors, those benefits must be considered realistically in light of the disclosures at issue and balanced against the benefits of those disclosures. Here, as in the context of gaming, carefully balancing these concerns requires considering not only whether to disclose, but also what information should be disclosed to whom and when. Where trade secrecy is not needed to incentivize innovation, it cannot justify non-disclosure that undermines accountability. Even where trade secret protection provides some innovation incentive, it will often be possible to design limited or staged disclosure regimes that provide enough exclusivity for recouping investments in innovation while still facilitating accountability and follow-on innovation. We discuss these possibilities in more detail below.

---

[60] Balafoutas & Kerschbamer, *supra* note 59, at 2.

[61] Keith Dowding & Brad R. Taylor, *Algorithmic Decision-Making, Agency Costs, and Institution-Based Trust*, 37 PHIL. & TECH. 68, 9–10 (2024); Warren J. von Eschenbach, *Transparency and the Black Box Problem: Why We Do Not Trust AI*, 34 PHIL. & TECH. 1607, 1618–19 (2021).

[62] *See* Selbst et al., *supra* note 4, at 470 ("If courts demonstrate a willingness to deconstruct [technology design choices], designers will have a greater incentive to design transparently to reduce litigation costs and will ultimately design more thoughtfully to incorporate not just efficiency and scaling considerations, but potential harms that they might have to litigate later."); Laufer et al., *supra* note 9, at 8.

[63] *See, e.g.*, Tyler Whittemore, *Beyond the Black Box: The Case for a Prize System to Encourage Artificial Intelligence Innovation*, 2024 MICH. ST. L. REV. 249, 299–306 (2024).

### *C. Technically Inscrutable Algorithms*

Much has been made about the fact that some machine learning and AI algorithms are technically inscrutable in that they are difficult, if not impossible, for human decision-makers or decision-subjects to understand compared to the logic employed by human decision-makers. These models are derived inductively by optimizing them to fit large datasets. They often incorporate large numbers of input variables or learned parameters. These algorithms can exhibit technical opacity because the optimization process often results in complicated and non-intuitive mathematical relationships between inputs and outputs.[64]

There are well-documented tradeoffs between performance, as measured by decision quality metrics, and explainability, if taken to require comprehensibility of the computational process that connects particular inputs to particular outputs.[65] These tradeoffs can be used to justify the use of more technically opaque models.[66] The extent to which performance is degraded by adopting more explainable models varies, however, and the loss of performance may often be small.[67] For example, "accuracy" in machine learning (as defined by the data science literature) is not always the best measure of good decision-making from a policy perspective.[68] Explainable models, by contrast, may better serve values such as accountability, equity, and sensitivity to evolving circumstances.

[64] Bambauer & Zarsky, *supra* note 25, at 27 ("[I]t is important not to overemphasize these efficiency-related problems brought on by gaming . . . the right inquiry is . . . whether a set of proxies that seems to be superior to other methods of decision-making can become inferior under conditions of gaming.").

[65] *See generally* Anna Nezvijevskaia et al., *The Accuracy Versus Interpretability Trade-Off in Fraud Detection Model*, 3 DATA & POL'Y 12, (2021). "Accuracy" is usually defined in terms of the number of mistakes the algorithm makes on data similar to the training data. Approaches to "explainability" in the technical literature tend to focus on explaining the details of the "rule" linking inputs to outputs.

[66] *See* Selbst & Barocas, *supra* note 4, at 1111; Lipton, *supra* note 9, at 21 ("[T]he short-term goal of building trust with doctors by developing transparent models might clash with the longer-term goal of improving health care. Be careful when giving up predictive power that the desire for transparency is justified and not simply a concession to institutional biases against new methods.").

A detailed understanding of an algorithm is not always required to further a given policy goal such as reining in a self-serving decision-maker. Data scientists deploy rigorous approaches to interpret the outputs of technically opaque models.[69] Moreover, disclosures of system objectives or training datasets enable meaningful scrutiny of decision-making systems.[70] Hence, the extent to which technical opacity precludes useful disclosure is exaggerated.

Important points about technical opacity for present purposes are: i) that decision-makers' choice of algorithm controls its extent; ii) that the relative difference in performance between opaque and explainable models is often small; and iii) that even for technically non-explainable algorithms, disclosures that are useful for accountability and other purposes are often feasible. Designing a decision-making algorithm and determining what to disclose each involve tradeoffs. Absent a disclosure mandate, these tradeoffs remain at the discretion of the decision-maker. While technical opacity has some social benefits, it may also enable decision-makers to avoid accountability. Thus, the situation is ripe for principal-agent problems, and claims about technical opacity should not be taken at face value.

---

[67] *See* Andrew Bell et al., *It's Just Not That Simple: An Empirical Study of the Accuracy-Explainability Trade-off in Machine Learning for Public Policy*, PROC. 2022 ACM CONF. ON FAIRNESS, ACCOUNTABILITY, & TRANSPARENCY 248, 255–57 (2022).

[68] *See generally* Robert Brauneis & Ellen P. Goodman, *Algorithmic Transparency for the Smart City*, 20 YALE J.L. & TECH. 103, 122–26 (2018).

[69] *Mapping the Mind of a Large Language Model*, ANTHROPIC (May 21, 2024), https://www.anthropic.com/news/mapping-mind-language-model [https://perma.cc/LXG7-XJ5N]; *Extracting Concepts from GPT-4*, OPENAI (June 6, 2024), https://openai.com/index/extracting-concepts-from-gpt-4/ [https://perma.cc/RDX4-9ZR6].

[70] Selbst & Barocas, *supra* note 4, at 1133–1135. *See also* Selbst et al., *supra* note 4, at 469 ("[I]t is imperative that courts start to deconstruct design choices in ML … technologists need to be far more transparent about the nature of the choices they make when designing new technology. They must create detailed documentation of the design choices made and the rationales for them."); Diakopoulos, *supra* note 26, at 199–204.

### *D. Social Benefits of Disclosure*

Assessing the purported social benefits of opacity in turn requires consideration of the social benefits of disclosure. This latter analysis may be characterized as "what is disclosed to whom?" As we noted above (and the patent system attests), disclosures about a decision-making system to potential competitors can promote innovation.[71] Meanwhile, disclosure to decision-subjects can facilitate socially beneficial activities, such as investing in education or better financial habits.[72]

Disclosure to decision-subjects (whether public or private) is important for perceptions of decision legitimacy. Disclosure to the public improves accountability with respect to decision quality, bias, and investment in decision system improvements.[73] We systematize the benefits of public disclosure into three categories: legal compliance, correction of errors and biases, and procedural rights linked to accountability.

First, disclosure about decision-making systems provides the information necessary to demonstrate and audit compliance with relevant regulations and policies. This information can include documentation of the data used, the processes followed, and the decisions made by the algorithms.[74] Disclosure thereby helps ensure that organizations are meeting specific regulatory requirements and internal policies.[75] As a result, it can incentivize the identification and mitigation of risks including data breaches, security vulnerabilities, and ethical concerns.[76]

[71] *See, e.g.*, Dyer et al., *supra* note 41 (finding that higher-quality disclosures of patented information result in spill-over effects that promote follow-on innovation); Kim & Valentine, *supra* note 41 (finding the same, despite proprietary costs on the disclosing firm); PASQUALE, *supra* note 4, at 153 ("There is little evidence that the inability to keep such systems secret would diminish innovation.").

[72] Diakopoulos, *supra* note 26, at 206; Kleinberg & Raghavan, *supra* note 10, at 19:2–6.

[73] Sandra Wachter & Brent Mittelstadt, *A Right to Reasonable Inferences: Re-Thinking Data Protection Law in the Age of Big Data and AI*, 2019 COLUM. BUS. L. REV. 494, 619–20 (2019). *See also* COFONE, *supra* note 14, at 38–44; Selbst et al., *supra* note 4, at 420–22.

[74] Stephen Casper et al., *Black-Box Access is Insufficient for Rigorous AI Audits*, PROC. 2024 ACM CONF. ON FAIRNESS, ACCOUNTABILITY, & TRANSPARENCY 2254, 2255–56 (2024); Karl Werder et al., *Establishing Data Provenance for Responsible Artificial Intelligence Systems*, 13 ACM TRANSACTIONS ON MGMT. INFO. SYS. 22, 22:8–10 (2022).

The second category of disclosure benefits includes error and bias correction. Disclosure helps identify discriminatory practices, whether intentional or unintentional, in automated systems.[77] Bias identification may also incentivize decision-makers to identify and address biases or discriminatory practices in their algorithms proactively, an important benefit in areas such as hiring, lending, and criminal justice.[78] Disclosure can also incentivize efforts to reduce errors and improve decision quality. Keeping a highly inaccurate proxy secret deprives decision-subjects of the opportunity to identify and complain about decision quality and disparate impacts that it may cause. Moreover, false negative and false positive decisions resulting from an inaccurate proxy are often distributed unevenly and, as a result, the burdens of flawed proxies are borne disproportionately by marginalized groups. These costs are reduced if the public has sufficient information to identify and demand correction of flawed proxies.

---

[75] *See* Calo & Citron, *supra* note 47, at 800 ("Challenging automated decisions was difficult because systems lacked audit trails that could help excavate the reason behind the decisions."); Balázs Bodó et al., *Tackling the Algorithmic Control Crisis—The Technical, Legal, and Ethical Challenges of Research into Algorithmic Agents*, 19 Yale J.L. & Tech. 133, 164–67 (2017).

[76] Citron & Pasquale, *supra* note 4, at 24–25 ("The FTC should be given access to credit-scoring systems and other scoring systems that unfairly harm consumers. Access could be more or less episodic depending on the extent of unfairness exhibited by the scoring system. Biannual audits would make sense for most scoring systems; more frequent monitoring would be necessary for those which had engaged in troubling conduct."); Burrell, *supra* note 34, at 10; Pasquale, *supra* note 4, at 145–47.

[77] Cynthia Dwork & Deirdre Mulligan, *It's Not Privacy, and It's Not Fair*, 66 Stan. L. Rev. Online 35, 36 (2013); Burrell, *supra* note 34, at 9 ("Finding ways to reveal something of the internal logic of an algorithm can address concerns about lack of 'fairness' and discriminatory effects, sometimes with reassuring evidence of the algorithm's objectivity."). *See generally* Nima Kordzadeh & Maryam Ghasemaghaei, *Algorithmic Bias: Review, Synthesis, and Future Research Directions*, 31 Eur. J. Info. Sys. 388 (2022); Finale Doshie-Velez & Been Kim, *Towards a Rigorous Science of Interpretable Machine Learning*, arXiv (Mar. 2, 2017), https://arxiv.org/abs/1702.08608 [https://perma.cc/5HSV-9ZGZ]; Moritz Hardt et al., *Equality of Opportunity in Supervised Learning*, arXiv (Oct. 7, 2016), https://arxiv.org/abs/1610.02413 [https://perma.cc/L24W-Y9WX]; Aaron Springer & Steve Whittaker, "*I had a solid theory before but it's falling apart": Polarizing Effects of Algorithmic Transparency*, arXiv (Nov. 6, 2018), https://arxiv.org/abs/1811.02163 [https://perma.cc/HEL2-7J8C]; Ferguson, *supra* note 3, at 53–60.

[78] Andrew Tutt, *An FDA for Algorithms*, 69 Admin. L. Rev. 83, 104–11 (2017). *See generally* Upol Ehsan et al., *Expanding Explainability: Towards Social Transparency in AI Systems*, CHI Conf. on Hum. Factors Comput. Sys. 1 (2021), https://doi.org/10.1145/3411764.3445188.

Third, disclosure enables explainability of automated systems and, therefore, accountability in the process of making and applying them.[79] With detailed information, decision-makers can better address questions from regulators, stakeholders, or affected individuals.[80] Explanation of government policy-making is a core requirement of procedural due process and an accountability mechanism.[81] Transparency also builds trust between decision-makers and decision-subjects.[82]

In sum, disclosure of the proxies and procedures used in decision-making confers significant social benefits by promoting accountability; improving decision performance; exposing (and thereby deterring) bias; and enabling decision-subjects to challenge the bases for erroneous decisions. Revealing information also incentivizes organizations to understand and validate how their automated systems make decisions, which helps ensure compliance with relevant regulations, mitigate risks, and build trust with stakeholders. Cumulatively, disclosure provides useful information to the commercial market regarding opportunities for socially beneficial improvement and innovation.

---

[79] Margot E. Kaminski & Jennifer M. Urban, *The Right to Contest AI*, 121 Colum. L. Rev. 1957, 2035 (2021) ("Individuals cannot correct inaccurate decisions if they cannot see the incorrect data, reasoning, or inferences underlying decisions. Individuals cannot be assured that decision-making is being applied nonarbitrarily if they cannot understand a decision-making system's logic."); Selbst & Barocas, *supra* note 4, at 1118–26. *See generally* Jake Goldenfein, *Algorithmic Transparency and Decision-Making Accountability: Thoughts for Buying Machine Learning Algorithms*, *in* Closer to the Machine: Technical, Social, and Legal Aspects of AI 41 (Cliff Bertram et al. eds., 2019); Reuben Binns et al., "*It's Reducing a Human Being to a Percentage": Perceptions of Justice in Algorithmic Decisions*, CHI Conf. on Hum. Factors Comput. Sys. 1 (2018), https://doi.org/10.1145/3173574.3173951.

[80] Calo & Citron, *supra* note 47, at 832 ("But agency officials do not appear to understand the [automated] systems they have commissioned to carry out this task. Crucially, they cannot explain them in public or in court because they do not know how they work."); Citron & Pasquale, *supra* note 4, at 31 ("Secret credit scoring can undermine the public good, since opaque methods of scoring make it difficult for those who feel-and quite possibly are-wronged to press their case."). *See also* Coglianese & Lehr, *supra* note 3, at 1205–13.

[81] Calo & Citron, *supra* note 47, at 800–803; Karen Levy et al., *Algorithms and Decision-Making in the Public Sector*, 17 Ann. Rev. L. & Soc. Sci. 309, 320–21 (2021). *See also* Margot E. Kaminski, *The Right to Explanation, Explained*, 34 Berkeley Tech. L.J. 189, 209–217 (2019); Kate Crawford & Jason Schultz, *Big Data and Due Process: Towards a Framework to Redress Predictive Privacy Harms*, 55 B.C. L. Rev. 93, 124–28 (2014).

[82] *See* von Eschenbach, *supra* note 61, at 1619–20; Bingjie Liu, *In AI We Trust? Effects of Agency Locus and Transparency on Uncertainty Reduction in Human–AI Interaction*, 26 J. Comput.-Mediated Commc'n 384, 386 (2021). *See generally* Ehsan et al., *supra* note 78; Marco Ribeiro et al., "*Why Should I Trust You?": Explaining the Predictions of Any Classifier*, Proc. 22nd ACM SIGKDD Int'l Conf. on Knowledge Discovery & Data Mining 1135 (2016). *Cf.* Mike Ananny & Kate Crawford, *Seeing Without Knowing: Limitations of the Transparency Ideal and Its Application to Algorithmic Accountability*, 20 New Media & Soc'y 973, 5–10 (2018).

### *E. Disclosure and its Trade-offs*

Disclosure can create social trade-offs in two distinct ways. First, disclosure to particular audiences might involve trade-offs. For example, decision-subjects might use information about how their characteristics affect a decision proxy either to game the proxy, engage in socially valuable error correction, or improve their true eligibility for a beneficial outcome. Disclosure to competitors, meanwhile, might lead either to free-riding that undercuts innovation or to valuable follow-on innovation and improvement. For present purposes, we assume that disclosure to "the public" or a public-appointed oversight body leads to socially desirable accountability, although ensuring that outcome represents its own complex governance issue. Second, there may be trade-offs when disclosure intended for one "audience" is used by another in the absence of a practical separation mechanism. Decision-subjects, members of the public, and competitors are overlapping stakeholders—an individual might belong to all three groups. Disclosing certain information about a decision proxy to "the public" might support accountability, while disclosing the same information to "decision-subjects" could lead to gaming and disclosing it to "competitors" could depress incentives for innovation. However, it is possible to mitigate this risk since different audiences may benefit most from the disclosure of *different information* and there may be mechanisms to separate them. For example, decision-makers may make disclosures to oversight bodies rather than to individual members of the public, who may themselves be decision-subjects. The bottom line is that asking only *whether to disclose* a decision-making system leads to a less helpful determination about disclosure than asking *what* to disclose, to *whom*, and *when*.

## II
## Decision-Makers' Principal-Agent Problem

Those who design or deploy decision-making systems should not always be trusted to design corresponding disclosure strategies.[83] They may fail to account for the social costs of bias or inaccuracy in designing decision proxies; have self-serving incentives to omit details and avoid accountability; and discount the social value of disclosure. Thus, algorithmic designers may produce socially sub-optimal proxies that either conceal the inadequacies of their designs or avoid competition. Self-serving justifications for hiding these behaviors include gaming and trade secrecy.

[83] Dowding & Taylor, *supra* note 61, at 16.

Principal-agent problems between those who design decision-making AI and the public operate at two sequential stages: at algorithm design and at resisting disclosure. First, divergences manifest at the design stage because nearly every decision-making process, whether automated or not, relies on some approximation of the ideal decision criteria.[84] We use the term proxy to denote this approximation in any form an AI system takes: a predicted score or classification, a recommendation or determination in generative systems, or a decision output in agentic systems. In each case, the system produces a stand-in for an ideal criterion that cannot be directly measured or fully specified in advance. The design of such proxies determines systematic patterns of error, bias, and cost allocation (i.e., how noisy or biased the decisions will be, as well as the private and social costs of the decision-making process itself). Decision-makers may fail to appropriately balance the social costs of inaccuracy and bias against the private costs of correcting those problems. Second, decision-makers often have self-serving incentives to resist disclosure to hide the details of their decision-making bases and procedures from those to whom they are accountable, such as supervisors, government officials, or the public at large.[85] We examine these in Sections A and B, respectively.

Two further reasons compound at both stages. First, because transparency mechanisms have positive externalities, decision-makers may not adequately account for the intrinsic value of disclosure to decision-subjects in allowing them to strategically correct errors and improve their real eligibility for beneficial decisions.[86] Second, decision-makers can give undue weight to the private benefits of trade secret exclusivity in comparison to the public benefits of competition and follow-on innovation. We examine these in Section C.

---

[84] *See* Ignacio Cofone & Warut Khern-am-nuai, *The Overstated Cost of AI Fairness in Criminal Justice*, 100 IND. L.J. 1431, 1458–1460 (2025).

[85] Lu, *supra* note 24, at 131–32.

[86] *See, e.g.*, Margot E. Kaminski & Jennifer M. Urban, *The Right to Contest AI*, 121 COLUM. L. REV. 1957, 2001 (2021) ("Rather than acting as mere 'abstract and vague' concepts, dignity and autonomy interests animate calls for accuracy (to root out harmful mistakes) and rule of law constraints (to root out unequal, inconsistent effects.")); Katherine J. Strandburg, *Rulemaking and Inscrutable Automated Decision Tools*, 119 COLUM. L. REV. 1851, 1885 (2019). *See also* Selbst & Barocas, *supra* note 4, at 1120.

## *A. Decision-Makers are Imperfect Agents of Society's Interests*

### *1. Both Public and Private Sector Decision-Makers Are Agents of the Public*

Both public agencies and private firms that make rights-affecting decisions through the design or implementation of algorithmic systems should be treated as agents of the public because they control operational choices that determine rights-affecting outcomes, they can advance or undermine the public's interests through those choices, and they act under conditions where outsiders cannot reliably observe effort, design trade-offs, and departures from standards. The public's welfare improves or worsens depending on how these actors set criteria and manage errors, yet the public typically sees only the outputs—not the underlying judgments, compromises, or strategic omissions that produced them. Framing this process only as externalities (where decision-makers might harm individuals through errors) misses what is distinctive: the failure is not merely that harms spill onto others, but that the party with authority over the process has informational advantages, creating incentives to shirk, distort, or conceal in ways the principal cannot easily detect or correct.

In democratic societies, public sector decision-makers are regularly and appropriately viewed as agents of the public because they exercise delegated authority on the public's behalf and within mandates set by law. But this dynamic is broader.

In private-sector domains such as employment, housing, education, and credit, decision-makers should also be treated as agents and the public as the principal because these firms are not merely pursuing private ends through ordinary market exchange; they are exercising power over individuals' lives in ways that law constrains in the name of collective values. Even though decisions are made by private actors, anti-discrimination law, consumer protection, and related regulatory regimes impose public-regarding duties that convert private decision-makers into the public's agents for certain purposes: they are tasked with implementing the public's will that decisions not be arbitrary, biased, or exclusionary in ways law deems unacceptable.

In numerous contexts, therefore, private sector decision-makers are tasked by regulation with responsibilities to carry out the public will by correcting for externalities, overcoming information asymmetries, and adapting to distributive concerns.[87] Regulations, such as those banning discrimination, reflect the public interest in the decisions of private actors, converting private actors into agents of the public for certain purposes.[88] More broadly, market economies are premised on the assumption that the pursuit of private economic interest will typically improve overall social welfare.

### 2. *Why Decision-Makers are Imperfect Agents*

Decision-makers' principal–agent problems arise at two levels. At the design stage, algorithmic decision-making relies on proxies for ideal criteria, and system designers and those who procure AI systems have incentives to select cheaper, more administratively convenient, or strategically advantageous proxies even when they impose social costs through systematic error or bias—costs that are often externalized onto decision-subjects and society rather than borne by the decision-maker. At the disclosure stage, disclosure imposes private costs (liability, reputational harm, loss of discretion, and competitive exposure) while it provides public benefits (accountability, contestability, deterrence of low-quality design, public trust). It is therefore unsurprising that decision-makers resist transparency to avoid scrutiny. It is not that decision-makers are bad actors, but that their incentives predictably diverge from the public interest, particularly in contexts where the social costs of inaccuracy and bias are externalized and where private actors lack the accountability infrastructure that constrains public agencies.

The fact that (public) decision-makers are imperfect agents of society's interests is hardly news. The public choice literature explains how government actors' private interests predictably distort their behavior away from the public interest they were appointed to and are meant to serve.[89] Government decision-makers may shirk their responsibilities to the public by, for example, investing less in decision-quality than would be socially optimal when low effort is difficult to detect from outcomes alone. They may overweight reputationally salient mistakes (e.g., releasing a defendant who reoffends) relative to less visible but pervasive harms (e.g., unnecessary detention). Explanation of government decisions as a core requirement of procedural due process is intended, at least in part, as an accountability mechanism to address these principal-agent problems.[90]

[87] *See, e.g.*, Citron & Pasquale, *supra* note 4, at 22 ("Given scoring's sensitivity, fair, accurate, and replicable use of data is critical. We cannot rely on companies themselves to 'self-regulate' toward this end-they are obligated merely to find the most efficient mode of processing, and not to vindicate other social values including fairness.").

[88] The appropriate scope of such duties is a matter of contentious debate, which we do not attempt to engage here.

[89] *See* DENNIS C. MUELLER, PUBLIC CHOICE III, 359–85 (2003).

Private sector actors tasked with public-regarding responsibilities are no less likely to behave strategically to avoid those responsibilities in service of private ends. The individual incentives to evade or dilute constraints are often even stronger in commercial settings in terms of rewards. The persistence of regulatory avoidance—and the enormous government resources devoted to auditing, enforcement, and litigation to deter it—reflects the baseline expectation that regulated entities will strategically adapt around constraints.

One kind of agent strategic behavior involves shirking: underinvestment in the principal's wellbeing when effort and quality are hard to observe. A designer or procurer may invest little effort in improving decision quality when it is hard for decision-subjects or the public at large to ascertain how difficult and costly it would be to do better. A designer may also invest sub-optimally in reducing particular sorts of errors for other reasons, including bias or private preferences. This kind of underinvestment is often not obvious to decision-subjects and the public (unless the resulting decision quality is abysmally low).

A second kind of distortion enabled by a principal-agent problem occurs when a designer or procurer over-invests in eliminating the types of errors most likely to be noticed or blamed on them or on the deployer while under-investing in eliminating others. Strategic considerations might also lead them to invest too much in other metrics that look good rather than interventions that matter most, for example by reducing the overall number of errors rather than focusing on the types of error reductions that provide the most social benefit because they are most harmful or inequitable.

Automating some or all of a decision-making process is sometimes intended (or at least presented) as a means for avoiding such principal-agent problems by removing self-interested humans from the loop.[91] But it does not eliminate the incentive problem. Automation instead relocates discretion, and therefore the strategic behavior, upstream to the point at which the automated process is designed, developed, or procured.[92] That is the lesson of the Amazon hiring story from the introduction: ostensibly to reduce historically gender-biased hiring practices, Amazon trained an algorithm to replace human decision-making but it did so with its historically gender-biased hiring data. Before, humans made individual gender-biased *hiring* decisions; after, humans made an *algorithmic design* decision that introduced gender bias. Automation was not a neutral process that removed discretion but, rather, it embedded choices about data, system design parameters (such as labels, features, or generation objectives), decision criteria, and procurement similarly subject to incentive misalignment.

[90] Levy et al., *supra* note 81, at 320–321. *See also* Kaminski, *supra* note 81, at 209–217; Crawford & Schultz, *supra* note 81, at 124–28.

### *B. Decision-Makers Can Use Secrecy Strategically*

The tension between society's interests and decision-makers' interests not only affects how decision-making systems are designed but also gives decision-makers incentives to avoid accountability by embracing opacity.[93] Secrecy can be used to cover up decision-maker lapses, biases, and shirking. Technical opacity, often called inscrutability, can provide additional "cover" for decision-maker refusals to disclose additional information that could be disclosed and understood, such as developer choices of system objectives, design parameters (e.g. features in predictive systems), input data, and so forth.

When opacity protects private benefits, decision-makers can be expected to overstate the risk that disclosure will degrade decision-making performance by allowing decision-subjects to game the system (as opposed to when disclosure would primarily enable error correction and legitimate eligibility improvement). They can likewise be expected to overstate the importance of trade secrecy for promoting that type of socially valuable innovation (as opposed to constituting the private benefit of avoiding competition and follow-on improvements).

[91] Dowding & Taylor, *supra* note 61, at 12–13.

[92] Coglianese, *supra* note 37, at 6–9; Strandburg, *supra* note 86, at 1862–63. *See also* Noble, *supra* note 2, at 90 ("The more we can make transparent the political dimensions of technology, the more we might be able to intervene in the spaces where algorithms are becoming a substitute for public policy debates over resource distribution—from mortgages to insurance to educational opportunities.").

[93] *See* Bambauer & Zarsky, *supra* note 25, at 47 ("Lawmakers should make their value hierarchies more transparent so that they can be challenged where the tradeoff does not match democratic expectations or common sense, and so that the law can develop in a more internally consistent way."). *See also* Tal Zarsky, *Transparent Predictions*, 2013 U. Ill. L. Rev. 1503, 1533 (2013) ("The most basic and popular justification for transparency is that it facilitates a check on governmental actions. These actions might be flawed, biased, ineffective, or inefficient. The relevant officials might be improperly balancing rights and interests, led by their own bigotry, or over-influenced by private interests.").

Because disclosure targets informational asymmetries, mandating it is a legal mechanism for mitigating principal-agent problems and aligning private and public-sector decision-making with public interests. For example, fair credit laws not only prohibit certain forms of discrimination but also demand a certain level of disclosure to applicants about the bases for loan denials. In other areas, such as employment and housing, the law does not require disclosure of decision-making criteria, but does directly prohibit reliance on certain characteristics, such as ethnicity, gender, age, and disability.

This is not to suggest that decision-makers are unconcerned with making sound decisions or that their warnings about the risks of disclosure should go unheeded. The point is that, when conflicting interests are at stake, decision-makers may not make socially optimal tradeoffs between investments in decision quality and the social costs of various sorts of errors and, as a result, may exaggerate the risk of disclosure to protect their private interests.[94]

In sum, both public and private decision-makers may engage in socially suboptimal strategic behavior. Undesirable strategic behavior is potentially more likely with private decision-makers because of different incentives and accountability structures in the two contexts.[95] Often, the social harms of inaccuracy and bias are externalized from decision-makers to society—and outsiders cannot easily observe whether these harms are avoidable, making accountability depend on transparency. Thus, when asking whether to mandate disclosure or to trust public or private entities to choose what or whether to disclose, one should ask: Do decision-makers have the right incentives to make the tradeoff? From a decision-maker's self-serving perspective, it may be preferable to strategically hide socially problematic details of the decision-making system rather than to design a better one.[96]

[94] PASQUALE, *supra* note 4, at 56.

[95] Balkin, *The Three Laws of Robotics*, *supra* note 9, at 1240 ("Rather, the goal, as more and more companies shift to algorithmic decision-making, and increase their levels of decision-making activity, is to require firms to adopt methods that are justified from the standpoint of society as a whole."). *See also* Kaminski, *supra* note 34, at 1397–98.

Because of these principal-agent problems, intentional secrecy can be problematic even when some aspects of a decision algorithm are technically opaque. Aspects of decision systems that could be disclosed but are deliberately hidden are more suspicious than technically inscrutable aspects. That said, intentional opacity can arise at all stages of design and implementation: decision system designers may strategically choose technically opaque systems to perform functions that transparent systems could perform nearly as well.

### *C. A Framework for Principal-Agent Problems in Algorithmic Decision-Making*

#### *1. Proxy Design and Accountability*

Detailed disclosure and auditing of decision-making systems is an effective way to avoid principal-agent problems at the proxy design tier. But what if there are potentially legitimate gaming or trade secrecy concerns? When should one suspect that principal-agent problems underlie claims about gaming or trade secrecy? Putting the above discussion together, one can categorize the contributors to the principal-agent problems associated with disclosure as follows:

One divergence of interests creates principal-agent problems at the level of proxy design. While the market will, to some extent, ordinarily reward algorithmic designers for developing high quality decision-making systems, designers do not fully internalize the social benefits of decision quality: many of these benefits operate as externalities. As a result, algorithmic designers may prefer to shirk, indulge their biases, or act in other self-serving ways when designing a proxy. Principal-agent problems at the proxy design level also generate a divergence of interests regarding disclosure to the public, since disclosure is an accountability mechanism. Decision-makers do not internalize the benefits of accountability, perceiving disclosure primarily in terms of private costs and penalties.[97] These private costs often include not only the need to invest more in decision quality (rather than shirking), but also the loss of (unwarranted) trust and perceptions of legitimacy.

[96] Graves & Katyal, *supra* note 6, at 1341 ("[C]ompanies learn that labeling sensitive or embarrassing information as a 'trade secret' or 'confidential' can stall or silence calls for disclosure. Further, as the reach of software in government agencies and decisionmaking increases apace, it becomes even more difficult to strategize for greater transparency in a world that increasingly relies upon automated, black-box decisionmaking.").

[97] *See* COFONE, *supra* note 14, at 59–66, 111–19.

*2. Analyzing Disclosure Given Principal-Agent Issues*

As discussed above,[98] decision-makers often have two types of concerns. These might not perfectly align with social concerns.

The first is strategic decision-subject behavior. Gaming is socially costly because it reduces decision quality, while strategic error correction and eligibility improvement improve decision quality. While decision-makers receive rewards for proxy quality in some circumstances, their private interests regarding strategic decision-subject behavior can diverge from those of society. For example, they do not fully internalize the social benefits of error correction and eligibility improvement; beneficial decisions can be privately costly if deployers are tasked with providing services to successful decision-subjects. Decision-makers in these contexts will tend to oppose disclosure more strongly than is warranted by the social costs of gaming.

The second is trade secrecy. While trade secrecy's social justification is that it maintains incentives for innovative investment, it also imposes social costs by preventing competition and making it harder to engage in follow-on innovation. Because commercial developers will internalize the benefit ore than the cost, they will overvalue the exclusivity provided by secrecy. This divergence between social and private interests creates a principal-agent problem.

The public's interest in "getting what it is paying for" weighs in favor of disclosing whatever information is necessary to assess decision quality and demand a socially optimal trade-off between quality and cost. Decision-makers may benefit from hiding their self-serving approaches to proxy design. While there is social value in avoiding disclosure that would facilitate gaming, decision-makers may undervalue the benefits of error correction and eligibility improvements. In a commercial context, decision-makers will nearly always overvalue secrecy without accounting for the social benefits of competition and follow-on innovation.

[98] *See supra* Sections I.A & I.B.

Moreover, decision-makers have strategic reasons to assert superficially plausible concerns about gaming and trade secrecy as justifications for resisting disclosure and avoiding accountability for suboptimal proxy design (one could hardly expect decision-makers to argue against disclosure on the grounds that they want to continue shirking or acting in biased ways). Because of these potential principal-agent problems, it is important to develop mechanisms for determining when decision-makers' claims about gaming and trade secrecy are red herrings.

The next two Parts suggest mechanisms for diagnosing principal-agent problems of the types delineated here and designing appropriate disclosure regimes when decision-makers are agents of the public.

## III
## Diagnosing Algorithmic Principal-Agent Problems

In this Part, we provide a framework for diagnosing principal-agent problems when decision-makers argue against disclosing elements of their algorithmic decision-making systems. We first argue that clues indicating the presence and nature of principal-agent problems at the tier of algorithm design may be found in the profile of different sorts of errors and in the distribution of decision outcomes across groups. In some situations, a proxy's error profile (i.e., the distribution of error rates) is also suggestive of the weight that should be given to gaming and innovation concerns in devising a disclosure regime. Moreover, such error patterns and outcome distributions can be disclosed and assessed routinely without raising worries about either trade secrecy or gaming. Disclosure of error profiles and outcome distributions in and of themselves can provide some accountability and innovation benefits. It thus can—and should—be mandatory for decision-making systems in which society has a significant interest. Focusing on error profiles is also one step toward moving beyond all-or-nothing debates, pitting the benefits of secrecy against the benefits of disclosure.

The article's diagnostic tool (the error profile, defined here in terms of false positive and false negative rates) draws primarily from predictive AI, where such metrics are most directly computable, because those are the models used so far for decision-making. Generative and agentic systems require analogous performance metrics—such as systematic directional biases, calibration disparities across groups, or audit-based accuracy assessments—that serve the same diagnostic function within the framework.

## A. *Decision Quality: A Framework for Discussion*

### 1. *Why Decision Quality?*

We focus on error profiles and the distribution of decision outcomes across groups as initial clues for diagnosing principal-agent problems that might affect proxy design for three reasons. First, there is widespread agreement that decision quality is important. Commonly cited social benefits of disclosure also focus largely on improving decision quality, whether through accountability or by enabling strategic error correction or eligibility improvements by decision-subjects.[99] Both gaming and trade secrecy, similarly, are cited to justify non-disclosure to ensure or improve decision quality. Gaming is deemed a problem precisely because of concerns that it will reduce decision quality by allowing decision-subjects to obtain undeserved beneficial outcomes (or avoid deserved negative outcomes).[100] The trade secrecy argument contends that secrecy is needed to protect socially valuable innovations.[101] In this context, that means either improved quality for equivalent cost or cost savings for equivalent quality. The trend toward inscrutable machine-learning decision algorithms is also purportedly driven by concerns with decision quality. Error profiles, if measured in terms of deviation from ideal decision criteria, are sensitive metrics of decision quality. Even when deviations from ideal decision criteria cannot be measured precisely, approximate error profiles can usefully suggest where principal-agent problems are likely and further inquiry is warranted.

Second, we show that preliminary disclosure of a profile of the *types* of mistakes that a decision-making system produces helps policymakers and courts assess whether principal-agent problems are likely to have infected proxy design. If the error profile raises suspicions, justifications can be demanded and further disclosure, perhaps in stages, can be required. Looking at error profiles, rather than at any single metric for decision quality, is important because a proxy that produces very low overall errors may still be socially suboptimal. Improving decision quality may be costly and those costs may or may not be worth bearing in a given context.[102] Moreover, a seemingly reasonable overall error rate can mask socially undesirable disparities between demographic groups if these error rates are unevenly distributed by disproportionately falling on a minority group who suffers a detriment for such uneven decision quality. For instance, a proxy with an overall 10% error rate could still treat groups differently if Group A experiences only a 97% accuracy rate while Group B experiences 65%. Aggregate metrics would obscure this disparity. The principal-agent issue for proxy design is whether tradeoffs are being made in accordance with social values rather than twisted to further a decision-maker's private goals. As we discuss below, error profiles provide useful insights into that question.

[99] Selbst & Barocas, *supra* note 4, at 1118–26; Selbst et al., *supra* note 4, at 420–22; Citron & Pasquale, *supra* note 4, at 31; Kaminski & Urban, *supra* note 79, at 2035; Burrell, *supra* note 34, at 9; Calo & Citron, *supra* note 47, at 832; Wachter & Mittelstadt, *supra* note 73; Dwork & Mulligan, *supra* note 77, at 37; Tutt, *supra* note 78, at 109–10; Binns et al., *supra* note 79, at 9–10.

[100] *See, e.g.*, Bambauer & Zarsky, *supra* note 25, at 10.

[101] *See, e.g.*, Cowgill & Tucker, *supra* note 35, at 23; New & Castro, *supra* note 35, at 12.

Third, as already mentioned, while error profiles can be highly informative, disclosing them poses no threat to legitimate concerns about gaming and trade secrecy. While companies may assert that disclosing their error profiles will give their competitors actionable insights into their secret innovations, policymakers and courts should reject such assertions.

### 2. *Defining an Error Profile*

For the purposes of this discussion, we define a simple error profile for binary yes/no decisions, which is the standard form in machine learning. For AI systems with non-binary or generated outputs, analogous performance metrics (such as systematic directional biases, disparate error rates across groups, or calibration gaps) serve the same diagnostic function. Consider a proxy used to make a binary yes/no decision about a pool of decision-subjects, such as the Amazon algorithm used to determine whether to call each applicant for a hiring interview. Assume for now that there are some means to assess whether this proxy yields the right decisions from a social perspective.[103] In this situation, decision outcomes are commonly described in terms of two binaries: positive and negative, and true and false.

[102] This point is well appreciated in case law, for example in the *Mathews* test for due process. *See* Mathews v. Eldridge, 424 U.S. 319 (1976).

[103] As a practical matter, such an assessment may of course be difficult, but one can imagine various mechanisms for assessing decision quality on a case-by-case or overall basis. This is, in fact, what is done on one end of the spectrum, by an appeals process, and on the other end, by the typical machine learning metric for accuracy.

Here, we call a decision "positive" if it is beneficial to the decision-subject and "negative" if it is not. Thus, in describing a decision as "positive" or "negative," we adopt the decision-subject's perspective.[104] In the hiring context, a decision to call an applicant in for a hiring interview is a "positive" and a decision not to do so is "negative," while in law enforcement, a decision to release a defendant pending trial is "positive" and a decision to keep them in jail is "negative." We also label decisions as "true" or "false," this time taking a social perspective. Thus, a decision is "true" if it is correct from a social perspective and "false" if it is mistaken from a social perspective. In the Amazon algorithm example, qualified candidates called for interviews were true positives, unqualified candidates called for interviews were false positives, qualified candidates not called for interviews were false negatives, and unqualified candidates not called for interviews were true negatives. False positives and false negatives are both errors that would contribute to the standard data science calculation of "accuracy," but they may have different implications for decision quality.

Table 1: An Illustration of True and False Positives and Negatives as Defined.

| | **Beneficial treatment** | **Detrimental treatment** |
|---|---|---|
| **Deserved** | True positive | True negative |
| **Undeserved** | False positive | False negative |

In these terms, a decision-subject who uses disclosed information about a decision-making algorithm to "game" the system changes what would have been a true negative into a false positive, thus decreasing decision quality. On the other hand, a decision-subject who strategically uses disclosure of a decision proxy to correct an error in the algorithm changes what would have been a false negative into a true positive, improving decision quality. A decision-subject might also use disclosure strategically to improve their qualifications, changing what would have been a true negative into a true positive. In employment, gaming might mean lying about one's work experience on one's resume, while error correction might mean removing indications of gender from the resume if the algorithmic proxy mistakenly treats them as indications of a lack of suitability for the job. Improving qualifications might mean taking a course about a software package required for the position. Gaming decreases decision quality, while error correction and qualification improvement are socially positive improvements to decision-making.

[104] In the literature, while the terminology is somewhat arbitrary, "positive" is often used to refer either to a "yes" decision or to a situation in which some condition holds. We adopt our definition because defining positive and negative from the decision-subject's perspective, while defining true and false from society's perspective is aligned with our concern about decision-subject strategic behavior.

When one speaks of an error profile, one usually has in mind a table like the one above with number or percentages of errors of each type. Ideally, the error profile would be measured with respect to the "right" decision from a social perspective, according to the ideal decision criteria. Decision-makers do not ordinarily have access to the "right" decision for each case; if they did, they would not need a proxy. There are, however, various techniques for approximating an error profile, depending on the situation. Sometimes, a sample of roughly "right" decisions can be obtained by performing a more intensive review of a set of cases. Retrospective review can be used to evaluate whether previous decisions turned out to be "right." At a minimum, for a machine learning algorithm, there is usually a set of training or test data from which an approximate error profile can be computed. It is not that the computation of error profiles is a trivial matter, but that it is a common and well-known problem that must be resolved somehow if a decision-making system is to be evaluated for quality.

With these definitions in mind, in Section B, we analyze how various types of error profiles are informative of the likelihood that principal-agent problems are infecting proxy design. In Section C, we consider what those error profiles say about whether trade secrecy provides a socially beneficial reason to avoid or limit disclosure. Section D turns to questions about strategic decision-subject behavior in light of error profiles. Section E summarizes what this analysis says about when disclosure mandates are likely to be necessary because decision-makers strategically withhold disclosure.

### *B. Error Profiles as Clues to Principal-Agent Problems in Proxy Design*

In this Section, we consider how several general patterns of errors can provide clues to the presence of principal-agent problems in proxy design. First, consider the implications of overall decision quality. We define a high-quality decision-making proxy as one which produces a low number of both false positives and false negatives, while a low-quality proxy produces a high number of both types of errors. By "high quality" here, we mean a situation in which rates of both sorts of error are "low," even if there is more of one sort of error than the other. Similarly, "low quality" decision-making systems, as discussed here, are systems in which rates of both sorts of error are high, even if one is higher than the others. The question of when both sorts of error rates are low enough to call a decision-making algorithm "high quality" (or so high that one would call the algorithm "low quality") is a judgment call that may vary by context and require both expert and political input. Here, we assume that, in a given context, there is some socially determined definition of when false negative and false positive rates are either both low enough (or both high enough) to call the decision-making algorithm high (or low) quality. We consider a proxy to be of "moderate" quality when overall rates of both false negatives and false positives are intermediate between high and low.

For moderate-quality proxies, we emphasize the importance of looking at asymmetries and imbalances between demographic groups because an initial appearance of moderate quality can be deceiving. We are particularly concerned about proxies for which overall error rates are higher for disfavored groups and proxies for which asymmetries between false positives and false negatives are differently distributed for different groups because these sorts of error profiles are highly likely to reflect principal-agent problems in proxy design.

### *1. High-Quality Proxies*

Consider a decision-making proxy that gives a positive result to nearly everyone who should (from a social perspective) get a positive outcome and a negative result to almost everyone who should (again, from a social perspective) get a negative outcome. Such a system would produce few false positives and few false negatives. For example, suppose an employment screening algorithm makes few mistakes when using performance on a math test as a proxy for identifying candidates for a job requiring math skills. The system's true positive rate and true negative rate are both high and it makes few mistakes of either sort. If this proxy's error rates are similarly low for members of protected categories or other groups of concern, it is what we term high quality.

A high-quality proxy suggests that there are unlikely to be principal-agent problems at the design level. It is possible, in principle, that the decision-maker is overly assiduous and the societal optimum would be more error-prone. But such a situation would seem to be a special case, given that decision-makers generally are resource-constrained and would have little to gain from being overly persnickety.

Thus, high quality across social groups is ordinarily a clue that there are no major principal-agent problems at the level of proxy design. Accountability concerns provide no major societal reason to disclose (or not to disclose) details of such a high-quality proxy. It is thus reasonable to allow questions about innovation and strategic decision-subject behavior, which are discussed below, to drive the disclosure regime.

*2. Low-Quality Proxies*

At the other end of the spectrum, consider a low-quality proxy for which both the false positive and false negative rates are high across demographic groups. Such a low-quality proxy would contribute marginally, at best, to making the right decision. In most contexts, the use of such a low-quality proxy should be considered a serious warning of likely principal-agent problems at the level of proxy design. At a minimum, the use of such a low-quality proxy requires justification.

There are some contexts in which employing a low-quality proxy is socially justified because the costs of employing a more accurate proxy would outweigh the total error costs incurred by society. For some decisions, a rough cut is all that is necessary, fair, and practical, given the expense associated with doing better. There are decisions for which lotteries or simple age cutoffs are employed for just this sort of reason. A rough proxy might be particularly appropriate when performing a "first cut" before engaging in a more costly in-depth review, perhaps in an employment or educational context.

In summary, a low-quality proxy usually indicates that principal-agent problems have infected proxy design. At a minimum, low quality requires justification. Where it cannot be justified, it would be socially preferable to impose an appropriate disclosure regime to expose the principal-agent problems and set the stage for improvement. As we discuss below,[105] trade secrecy and gaming concerns are unlikely to justify keeping the details of a low-quality proxy secret.

*3. Symmetric Moderate Quality Proxies without Disparities between Groups*

Because it is often costly to improve a proxy's quality, the socially preferred decision proxy, in some situations, will have a moderate quality error profile that reflects a balance of the social costs and benefits of improving decision quality. For now, we consider moderate quality proxies without significant disparities between demographic groups or asymmetries between error types because, as we discuss below, such disparities and asymmetries often suggest principal-agent problems.

[105] *See infra* Sections III.C, III.D.

For a symmetric, balanced, moderate-quality proxy, the primary concern about proxy design is whether the designer or procurer is shirking or failing to invest sufficiently in decision quality for some other reason. When a proxy is used to make decisions about members of dominant social groups (i.e., populations with greater power, social standing, or institutional influence in the decision context), one might be less concerned, expecting that politics or market forces would counter this sort of slacking, although even dominant groups can be hoodwinked by claims about technical impossibility. When a decision-making proxy primarily affects decision-subjects from less influential social groups, however, shirking is more likely to go unredressed and failure to invest in improving the proxy may even reflect decision-maker biases. Many government decision-making contexts, from awarding benefits to imposing pre-trial detention, primarily affect decision-subjects from less powerful social groups. In such situations, even when decision quality is moderate and there are no obvious disparities in the error profiles, it would be socially desirable to check for principal-agent problems by imposing an appropriate disclosure regime.

#### *4. Moderate-Quality Proxies with Asymmetric Error Profiles*

If errors occur randomly, we expect the error profile to be symmetric. A moderate-quality asymmetric error profile suggests that a decision-maker either made a choice to favor one sort of error over the other or did not consider the implications of the asymmetry. (For high-quality proxies, asymmetries are relatively insignificant and unlikely to indicate serious principal-agent problems. Low-quality proxies usually suffer from serious principal-agent problems, regardless of asymmetry.) The question of whether the asymmetry indicates a principal-agent problem centers on whether there is a social justification for the asymmetry.

In some decision-making contexts, a degree of asymmetry is justifiable—and even desirable—because of the varying social costs of different types of errors or the varying investments needed to reduce them. A well-known example of a justifiable asymmetry results from the tradeoff between ensuring that guilty people are convicted and ensuring that innocent people are not. Because the burden of false imprisonment is extremely high, criminal law (at least rhetorically) takes the view that it is better to have ten guilty people go free than to punish one innocent person. It would be even better to have a high-quality proxy that makes few errors of either sort, but tradeoffs are often inevitable because improving decisions is costly and resources are limited. If there are only sufficient resources to create a moderate-quality proxy, the social preference for criminal trials would be to have a large (ten-fold) asymmetry between false positives and false negatives. Similarly, the social cost of denying food stamps to a family that needs them is almost certainly higher than the social cost of providing them to a family that does not. If it is too difficult or costly to devise a high-quality proxy to determine who should receive food stamps, a moderate-quality proxy with an excess of false positives over false negatives would be socially preferable. A social preference for an asymmetric error profile can also arise from asymmetries in the costs of eliminating different types of errors. If two types of errors are equally socially costly, but one is more difficult to avoid, the socially optimal approach would be to eliminate more of the errors that are easier to avoid.

These complexities leave substantial room for decision-makers to cover up principal-agent problems in proxy design. Consider, for example, a decision-making proxy that leads to far more false negatives than false positives. This means that the decision-maker is careful not to award benefits erroneously, but less careful about erroneously denying benefits. This may be the socially preferred outcome. But this sort of asymmetry could also occur for several types of self-serving reasons. Perhaps false positives are more visible than false negatives to those overseeing decision-maker performance and the decision-maker is risk averse. Perhaps positive outcomes impose private costs on a deployer (for example, they might have to provide services to decision-subjects) while the social benefits that outweigh those costs are externalized to decision-subjects or society at large. Conversely, a decision-maker who expects to have to cope with many complaints about negative decisions might have strategic reasons to skew the algorithm in favor of false positives. Or perhaps reducing false negatives is easier or cheaper for the decision-maker than reducing false positives (or vice versa). While such cost differentials can justify asymmetry, decision-makers might prefer asymmetry even when society would benefit from equalizing the error profile.

Even when a degree of asymmetry is socially desirable, decision-makers may have selfish reasons to skew the error profile away from the socially desirable point.

For example, a risk-averse lender might develop an algorithm for calculating credit scores that denies credit to many people who would have repaid to zealously avoid making loans to people who default. The resulting asymmetry might be socially appropriate for some credit determinations, but not for others. For example, society might benefit from a more risk-taking approach to educational loans considering the large and socially beneficial externalities of education. Conversely, a lender might structure transactions to benefit from a socially sub-optimal highly risk-taking approach, as was the case during the subprime mortgage crisis.

Asymmetries in the error profile might reflect appropriate accounting for normative concerns and balancing of social costs and benefits. But they might also result from socially undesirable behavior by the decision-maker—either in strategically preferring one sort of error to the other without social justification or in shirking socially valuable efforts to reduce the asymmetry. Thus, when there is no socially justifiable reason to have an asymmetric error profile, principal-agent problems should be suspected. The clue is not conclusive—the asymmetry may be an appropriate reflection of social values—but it is reason to push for justification of the asymmetry and for whatever disclosure is necessary to evaluate the proffered justification. Disclosure puts the public (or regulators) in a position to ask questions and demand accountability.

Even when a detailed assessment of the socially optimal degree of asymmetry would require specialized expertise and information about the context and the costs of reducing different sorts of errors, members of the public and various oversight groups will often have a rough idea of what the balance should look like. For example, the public would know something was seriously awry in the criminal justice system if the false negative rate outweighed the false positive rate. This rough sense that an asymmetry is problematic is an impetus to demand justification from decision-makers. In contexts where there is no obvious justification for a significant asymmetry, the public should be suspicious if the decision-maker is unable to provide a sensible explanation.

Thus, decision-makers should ordinarily be required to justify error profile asymmetries (or deviations from a socially appropriate level of asymmetry) and to disclose enough information about the proxy design to allow the public (or perhaps an oversight body) to determine whether principal-agent problems have occurred.

### 5. *Moderate Proxies with Error Profile Disparities Across Groups*

Until now, we have considered situations in which a decision proxy's quality is consistent across social groups (even though the proxy might, in practice, be applied primarily to one social group). However, one particularly weighty concern about opaque proxies is that they might allow decision-makers to hide bias, whether it is intentional or emerges from a given proxy design process. For example, a biased proxy can result unintentionally from using biased data to train a machine learning model, as likely occurred in the Amazon hiring example. Because bias means some decisions are mistaken in light of the ideal decision criteria, bias will be reflected in differences between the error profiles for different groups. To put it another way, if the error profile of two groups is the same, it is reasonable to conclude that the associated decision proxy is unbiased. Disparities between error profiles are highly suggestive of principal-agent problems.

There are two relevant types of error profile disparity between social groups: differences in overall error rate, which we call "overall disparity," and differences in asymmetries between false positives and false negatives, which we call "disparate asymmetry." In this section, we first consider overall disparities and then disparate asymmetries.

#### 5.1. *Overall Error Rate Disparities*

An overall error rate disparity between groups is often, though not always, an indicator of principal-agent problems. Sometimes, a disparity in error rates between groups is costly or difficult for decision-makers to avoid (for example, if the proxy is data-driven and there is less data available for one group than for another). Whether the disparity indicates principal-agent problems depends on whether it is socially desirable to invest enough in improving the proxy to eliminate the disparity. That, in turn, depends on the nature of the disparity and the decision context. If a disparity in decision quality is socially costly to remedy, there is sometimes a normative justification for tolerating a degree of overall disparity in error rates.

Nonetheless, overall disparities are always disadvantageous to the group receiving lower quality decisions, socially costly, and should raise suspicions about principal-agent problems. If the errors are symmetric between false positives and false negatives, it might superficially seem that the group ramifications of lower decision quality are a wash. That is often not the case. Beyond the individual unfairness to decision-subjects who receive false negative decisions, pernicious effects at the group level are possible. For example, there is empirical evidence that credit ratings are systematically noisier for individuals in certain social groups. [106] In this context, while an individual receiving a false positive decision may be happy in the near term, over the long term a proxy with high error rates for a group is likely to have a negative disparate impact. Not only are group members who receive false negative denials of credit deprived of the benefits of loans that they might have used to invest in businesses or otherwise improve their futures, group members who receive false positive awards of credit are likely to fail in the long run, driving themselves into worse financial straits. These mistakes are not only bad for these individuals. In the long run, these individual failures may rebound on other members of the group by reinforcing stereotypes of financial unreliability and, more directly, by providing negative data that may be fed back into credit scoring algorithms. Because many social groups are connected in relatively close social networks, failures to obtain deserved credit and failures to pay back unwisely awarded credit are also likely to have ripple effects on the financial health of other members of the group.

Because of these disadvantages, one should be most suspicious of principal-agent problems in proxy design when dominant populations receive higher quality decisions than disfavored groups. If a proxy makes significantly more errors for the dominant group (and the error profile is reasonably symmetric), one would ordinarily expect there to be political and market pressure for improvement. Members of dominant groups who receive false negative decisions are likely to complain, appeal, or otherwise make things difficult for the decision-maker. A disfavored group will have less political and market clout and often fewer resources to invest in seeking redress for erroneous decisions, making it more likely that low decision quality will go unaddressed. Moreover, when the socially dominant group outnumbers the disfavored group, overall moderate error rates can mask the low-quality results for the disfavored group. For these reasons, we focus our analysis on overall disparities that favor the dominant group.

[106] *How Flawed Data Aggravates Inequality in Credit*, Stan. Hum.-Centered A.I. (2023), https://hai.stanford.edu/news/how-flawed-data-aggravates-inequality-credit [https://perma.cc/JJ5L-QD3V]; Laura Blattner & Scott Nelson, *How Costly is Noise? Data and Disparities in Consumer Credit,* arXiv (May 17, 2021), https://arxiv.org/abs/2105.07554 [https://perma.cc/FFC4-G7KS]; Will Dobbie et al., *Measuring Bias in Consumer Lending*, 88 Rev. Econ. Stud. 2799, 2802, 2820 (2021).

Decision-makers have strategic incentives to tolerate such disparities. They may neglect to correct disparities between groups because they are focused strategically on investing in improvements for dominant groups. They may also shirk by investing less in improving the proxy for disfavored groups because they assume (perhaps correctly) that it will be more difficult or costly to develop higher-quality proxies for disfavored groups. For AI systems trained on historical data, relevant data about those groups is often less readily available, perhaps because members of the disfavored group were historically excluded (as in the Amazon hiring example), because the design parameters of the proxy have been chosen with the dominant group in mind, or because the dominant group is larger.[107] Disfavored groups may also be neglected because of intentional discrimination or implicit bias.

While designers and procurers bear the private costs of investing in improving proxy quality for socially disfavored groups, they often do not internalize many of the social benefits of such improvements. Such groups may also lack the clout to punish decision-makers for the lower-quality decisions they receive.

Overall disparities, especially if they favor the dominant group, are clues to principal-agent problems in proxy design. When such disparities are observed, disclosure is socially desirable to facilitate democratically accountable determinations about whether to tolerate the disparity or invest in creating a more evenhanded decision-making proxy. If, for example, society is dedicated to treating all decision-subjects equally but improving decision quality for one group is much more expensive than improving it for others, there is a normative decision to be made about how many resources to devote to equalizing outcomes considering the size of the disparity, the importance of the decision and other factors. This sort of moral and political determination ordinarily should not be made by decision-makers, who will have incentives to minimize private costs.[108] Thus, a mandatory disclosure regime will usually be necessary to provide accountability.

---

[107] Katja Langenbucher, *Responsible A.I.-based Credit Scoring–A Legal Framework*, 31 EUR. BUS. L. REV. 527, 554–56 (2020); Mikella Hurley & Julius Adebayo, *Credit Scoring in the Era of Big Data*, 18 YALE J.L. & TECH. 148, 202 (2016) ("The credit scoring industry is . . . relying on opaque scoring tools that use numerous data sources and proprietary algorithms in order to determine which consumers get access to credit . . . . [T]hese tools may combine facially neutral data points and treat them as proxies for immutable features such as race, thereby circumventing existing non-discrimination laws and denying credit access to certain groups.").

[108] *See* COFONE, *supra* note 14, at 59–66.

### 5.2. *Disparate Asymmetry*

Even if a proxy has similar overall error rates for two social groups, disparate asymmetries in the groups' error profiles are socially concerning. For example, a group that benefits from many false positives might receive the same total rate of erroneous decisions as a group that is penalized by a large number of false negatives. This was, in fact, the situation that led many to characterize the COMPAS recidivism prediction algorithm as racially biased. While the overall error rates were similar for black and white defendants, the distribution of those errors was racially skewed: Black defendants were more likely to be penalized by erroneous predictions of recidivism (false negatives in our parlance), while white defendants were more likely to benefit from erroneous predictions of non-recidivism (false positives in our terms).[109]

Disparate asymmetries can result from strategic neglect of disfavored groups. Decision-makers may be biased against the disfavored group (explicitly or implicitly) or expect dominant groups to have more influence on their performance reviews and private rewards. Decision-makers might strategically ignore false positives that benefit socially dominant groups (for example, a high rate of mistaken predictions that white defendants will not commit crimes if released pending trial), but are unlikely to strategically employ a decision system with high rates of false negatives for socially dominant groups (such as a high rate of unnecessary pretrial detentions for white defendants). For example, while studies suggest that drug use is similarly prevalent among individuals of various ethnicities,[110] enforcement is heavily skewed toward African Americans, meaning that the rate of false negatives regarding other racial groups is disproportionately high.

[109] This was the case, even setting aside the bias imported into the algorithm by the use of rearrest, rather than conviction, as the outcome proxy for recidivism despite the well-known effect of racial bias on likelihood of arrest. *See* Cofone & Khern-am-nuai, *supra* note 84, at 1447, 1457–61.

[110] Substance Abuse and Mental Health Services Administration (SAMHSA), 2021 National Survey on Drug Use and Health: Detailed Tables, Table 1.23B (2022), https://www.samhsa.gov/data/sites/default/files/reports/rpt39441/NSDUHDetailedTabs2021/NSDUHDetailedTabs2021/NSDUHDetailedTabs2021.htm#tab1-23b (on file with the NYU Journal of Intellectual Property and Entertainment Law); *2022 National Survey on Drug Use and Health: Race/Ethnicity Highlights*, SAMHSA (2023), https://www.samhsa.gov/data/sites/default/files/reports/rpt42731/2022-nsduh-race-eth-highlights.pdf [https://perma.cc/C8FS-8WFC].

Conversely, some algorithms used by the police (as did previously some investigation tactics such as "stop-and-frisk") have astonishingly high rates of false positives, making them both discriminatory and of limited use.[111] Predictive policing algorithm PredPol, for example, has been found to have 99.4% to 99.9% false positives depending on the type of crime.[112] Those high numbers of false-positives exist due to principal-agent problems. Police departments that purchase PredPol will complain if the system misses a crime, but not if it wrongly flags one. People wrongly flagged by the system do not have a say. The error rates are unevenly distributed, so marginalized populations get further marginalized. The outcome is not socially optimal, but it is privately optimal from the perspective of the decision-makers. The overall error rate for such a proxy—and even the error rate for each group separately—might be moderate. But from a social view, this sort of disparate asymmetry between groups is difficult to justify because it suggests bias and discrimination problems.

Just as for disparities in overall error rates, one source of disparate asymmetry is data availability. For example, in the child welfare arena, asymmetric imbalanced proxies result because algorithms used to detect child abuse and neglect rely on types of data that are systematically more available for economically disadvantaged groups. In this case, lack of data about the dominant group leads to an overabundance of false predictions of non-abuse (and a correspondingly low number of false predictions of abuse). Further, the data issue is potentially exacerbated by strategic concerns about the potential fall-out for the deploying agency if a member of the dominant group is falsely accused of child abuse. For the disfavored group, while one might expect the availability of more data to decrease errors of all types, natural concerns about neglecting vulnerable children lead to a risk-averse approach that produces a relatively large number of false predictions of abuse. Some asymmetry is probably socially preferable, given the severe consequences of abuse for vulnerable children. Nonetheless, the consequences of false predictions of abuse and neglect for families and children may also be severe. The social costs of those errors may not be internalized by proxy designers, especially when they fall heavily on disfavored social groups who are less likely to have effective avenues for redress and complaint.

---

[111] William Heisel, *It's Time to Address Facial Recognition: The Most Troubling Law Enforcement AI Tool*, BULL. ATOMIC SCI. (Nov. 10, 2021), https://thebulletin.org/2021/11/its-time-to-address-facial-recognition-the-most-troubling-law-enforcement-ai-tool/ [https://perma.cc/D9VH-Q8KC]; *When Artificial Intelligence Gets It Wrong*, INNOCENCE PROJECT (Sep. 19, 2023), https://innocenceproject.org/when-artificial-intelligence-gets-it-wrong/ [https://perma.cc/9W2P-ZCEQ].

[112] Aaron Sankin, *Predictive Policing Software Terrible at Predicting Crimes*, WIRED (Oct. 2, 2023), https://www.wired.com/story/plainfield-geolitica-crime-predictions/ [https://perma.cc/7BJ6-VYTN].

Consider also a situation, such as employment or college admissions, in which the number of “positive” slots (beneficial from a decision-subject’s perspective) is limited and a deployer employs a two-stage screening process to cut down on decision costs. If the number of potentially competent workers or good students exceeds available slots, the cost of a high rate of false negatives at the first stage may be negligible to deployers. In contrast, false positives might impose a significant burden on employers, necessitating costly additional screening. Employers and colleges are unlikely to be satisfied with a proxy that lets a high number of unqualified individuals through, even if it rarely screens out qualified candidates, because this sort of proxy forces them to conduct expensive interviews or employ other costly measures to further whittle down the pool.

While some asymmetric overall preference for false negatives is socially reasonable in this sort of situation to cut down screening costs, it can lead to disparate asymmetry if carelessly handled. Recall the Amazon hiring example. Most likely, the company was seeking candidates for a limited number of positions. From the company’s perspective, using historical employee data provided a ”safe” way to minimize false positives in the initial screening. If the company anticipates that there will be many reasonably qualified applicants, it may not be worried about false negatives. Society also benefits from avoiding unnecessary second-round screening costs. The principal-agent problem arises when the first round of screening is biased in socially undesirable ways. For example, if the company has better information for screening male applicants accurately, it can reduce false positives by rejecting more female applicants (even if not intentionally). This may produce a situation in which the error profile for male candidates has few false positives and more false negatives, reflecting a desirable overall asymmetry, but the error profile for female candidates has even fewer false positives and many false negatives. If there are plenty of qualified applicants, the company is not burdened by the social costs of the disparity in false negative rates associated with rejected female applicants who would have performed well.

In sum, disparate asymmetry in an error profile strongly suggests that there are principal-agent problems in proxy design. Disclosure would provide accountability in such cases.

### *C. Trade Secrecy as a Rationale for Non-Disclosure*

Trade secrecy is commonly invoked as a rationale for non-disclosure of decision proxies with the justification that it is required to incentivize innovative improvements. From a social perspective, worthwhile innovation in the context of a decision proxy means either that the proxy does a better job of reflecting the socially ideal decision criteria for the decision or that it lowers costs for a given decision quality. As discussed earlier, trade secrecy is generally a blunt tool for promoting innovation and is particularly so when it comes to credence goods such as decision-making proxies. Trade secrecy can also be selfishly invoked by decision-makers to resist disclosure, either to hide principal-agent issues in proxy design or to allow them to recoup profits in excess of what is necessary to recoup their investments.

#### *1. High Quality Proxies*

In commercial contexts, high-quality proxies provide the strongest justification for trade secrecy. High-quality proxies are unlikely to reflect principal-agent problems with proxy design and will often reflect substantial investment. If details about a high-quality decision-making algorithm are disclosed, competitors may be tempted to copy it and, especially because it is so good, may have relatively little incentive to invest in improving it further. Improving an excellent proxy is likely to be harder and more expensive than improving a low-quality proxy.

Even high quality is not necessarily a justification for unlimited trade secrecy, since there is always social value in disclosing information that can promote downstream innovation, perhaps aimed at reducing the cost of obtaining high-quality decisions. Even decision-makers who have faithfully designed high-quality proxies may strategically demand more secrecy than is necessary to recoup their investments because they want to maximize commercial profits (especially when their competitive advantage lies in something other than their ability to select accurately). Whether it is worthwhile to require additional disclosure for a high-quality proxy will depend on the context.

### *2. Low-Quality Proxies*

Incentivizing innovation provides little reason to keep details about low-quality proxies secret, since it is unlikely that low-quality proxies reflect substantial investment in socially valuable innovation. Disclosing details about a low-quality proxy is socially beneficial because it may give competitors ideas about how to improve things (potentially solving a market for lemons problem). Because low-quality proxies are not particularly innovative or valuable and presumably also not very costly to create, it would be almost perverse to worry about keeping them secret to provide incentives to create them. In sum, promoting innovation is not a reason to keep low-quality proxies secret. This sets up a significant principal-agent problem since commercial developers will nearly always prefer secrecy.

### *3. Moderate Quality Proxies*

As discussed above, while a moderate error profile may be optimal considering the costs of improving decision quality, a profile with disparities between social groups is unlikely to be socially preferable. Especially when the error profile is disadvantageous to a disfavored social group, disclosure to potential follow-on innovators should be prioritized over the potential innovation benefits of secrecy. Rather than keep secrets that prevent public assessment and competitive improvement, policymakers should focus on tailoring disclosure regimes to allow the oversight needed to ensure that innovation is aimed at the most socially valuable improvements, while providing alternative mechanisms for recouping innovative investments if necessary. Because commercial proxy developers do not internalize the benefits of disclosure, their assertions that unlimited trade secrecy is necessary to incentivize innovation should be viewed with skepticism even for moderate-quality proxies.

## *D. Strategic Decision-Subject Behavior and Disclosure*

Sometimes, the argument against disclosure is that it will facilitate socially undesirable strategic gaming by decision-subjects. Here, we discuss what can be learned from the error profile about the social importance of this concern. As discussed above,[113] disclosure of details about the decision proxy sometimes facilitates strategic behavior by decision-subjects. Gaming is socially detrimental because it changes true negatives to false positives, while strategic error correction and improvements to true eligibility for beneficial decisions are socially desirable. A decision proxy's error profile can give clues about the relative likelihood of these possibilities.

[113] *See infra* Section I.A.

### *1. High Quality Proxies*

High-quality proxies ordinarily are not easily gameable because of the close correlation between the proxy and the ideal decision-making criteria: it is hard to game a proxy that rarely makes mistakes.[114] Thus, for high-quality proxies, the threat of gaming ordinarily provides little justification for keeping details about the proxy secret.

However, there is a narrow category of cases for which disclosure might make it possible to game a high-quality proxy. Most of the time, the strength of the correlation between a high-quality proxy and the ideal decision criteria has an effective causal element that is beyond the control of decision-subjects. People who do well on a test that does a good job of examining the skills that will be required for a job are likely to do well in the job (low false positives), while people who do not have those skills are unlikely to do well on the test (low false negatives). Some applicants might panic or be sick and thus perform below their skill level on a skill test, but for a high-quality proxy, there are few such applicants.

Now consider another example: a proxy that uses volunteering for a charitable organization to screen for employment at an NGO that values devotion to community service. The fidelity of this proxy hinges upon the specific definition of "volunteering for a charitable organization." An algorithm that uses 5 years of volunteering for 10 hours per week as a proxy is likely to be much higher quality than an algorithm that uses a single stint of volunteering for 10 hours, because people who do not have the devotion to community service that the NGO job demands are unlikely to maintain the higher level of charitable engagement.

[114] Cofone & Strandburg, *supra* note 7, at 662 ("[D]ecision makers can respond to the threat of gaming by devising stronger proxies, thus simultaneously improving decision performance and making gaming more difficult. By adopting and disclosing more accurate proxies, decision makers can sometimes encourage decision subjects to invest in developing features that improve their qualifications for positive decision outcomes, often simultaneously producing better results for decision makers.").

This high-quality proxy would be hard to game for the same reason that it is so accurate: spending so much time on charitable work is costly, especially for those who are not devoted to community service.[115] Devotion to community service is (usually) the reason for doing a lot of volunteering for a charitable organization over a long period. Even if this proxy is disclosed publicly, it is unlikely that many potential applicants who are not committed to community service will invest five years in charitable work to improve their chances of obtaining an NGO job that requires a commitment to community service.

Next, consider the second version of the NGO's hiring algorithm. Because it is relatively cheap to put in a single stint of a few hours volunteering, this proxy is likely to be quite gameable. But such a proxy is also unlikely to be of high quality. There are many reasons why a person might spend 10 hours volunteering for a charitable organization, including trying to impress someone or beef up their reputation, fulfilling requirements of a community service alternative sentencing program, getting into a good college, and so forth. Most of those reasons are irrelevant to the NGO job, which requires a serious commitment to community service. Thus, a hiring algorithm that uses a brief stint of volunteering as a proxy will almost certainly have quite a few false positives. It is thus unlikely that such a proxy would be high quality.

As the example illustrates, a high-quality proxy will rarely be gameable because the low error rate usually reflects a causal element, while an easily gameable algorithm is unlikely to be high quality. However, there are some situations in which disclosure allows decision-subjects to "break" a previously tight connection between the proxy and the ideal decision criteria. Consider, for example, a "hideout" used exclusively by members of a criminal gang. In that case, being seen entering or exiting the hideout is a high-quality proxy for membership in the gang. Suppose police discover the location of the hideout and plan to raid it and arrest everyone there. If members of the gang discover that the police know the location of the hideout, they could game the proxy by moving their operations to a different location. Or they could degrade its quality by inviting non-gang members to parties at the hideout at random times. If the gang knows that police are using entering or exiting the hideout as a proxy for gang membership, they can game the proxy and destroy its validity. They are only able to do this, however, because the connection between the hideout and gang membership is under the control of the gang members and could be broken by them at will. Note, though, that breaking the connection between a high-quality proxy and the ideal decision criteria may require a potentially costly coordinated (or at least collective) change of behavior by decision-subjects.[116]

[115] The effort required to game high-quality proxies may also lessen concerns about self-surveillance by decision subjects. *See* Jane R. Bambauer et al., *When a Small Change Makes a Big Difference: Algorithmic Fairness among Similar Individuals*, 55 U.C. Davis L. Rev. 2337, 2385 (2022) ("If a system can be manipulated by changing behaviors, individuals will feel some amount of pressure to constantly monitor their choices and behaviors to optimize how they will be treated by an algorithm," especially in systems where "an individual can constantly change their conduct without too much cost at any given moment").

Gaming always reduces the validity of a proxy—that is both the problem with it and its purpose. The point here is that in most situations, low error rates reflect the sort of connection between proxy and ideal decision criteria that is costly or impossible for decision-subjects to break because it is hard to have one without the other. For a proxy to be both high quality and gameable, that correlation must be both nearly perfect and incidental, like the choice of a meeting place.

The upshot of the gameability analysis is that it is reasonable to presume that a high-quality decision proxy can be disclosed without worrying about gaming unless the decision-maker can explain why the connection between the proxy and the ideal decision criteria is unusually fragile.

On the flip side, high-quality proxies also provide little opportunity for strategic error correction (turning false negatives into true positives), if only because there are so few false negatives in the first place. Thus, strategic error correction does not provide a socially significant reason for disclosure.

Disclosure of a high-quality proxy might also provide a socially valuable opportunity for decision-subjects to improve their true eligibility for beneficial decisions. As discussed earlier, decision-makers might strategically resist such disclosures if favorable decisions are privately costly. The high quality of the proxy suggests this sort of principal-agent problem is unlikely, however. A deployer who preferred to make fewer favorable decisions would have procured a proxy skewed toward false negatives, or a designer serving such a client would have built one accordingly in the first place, regardless of whether decision-subjects were given the information needed to strategically improve their qualifications. It would be oddly coincidental if their tolerance for beneficial decisions ran out just for decision-subjects who could improve their eligibility after disclosure.

[116] Also relevant to the example is that this is not the final decision. The police may use the proxy of the hideout as the decision to investigate further, but the evidence used against the gang members in court is unlikely to include their drinking location, precisely because it is at best contextual. Therefore, factors that are relevant are that the proxy must be: (a) a high quality but random proxy (b) used in a final decision.

The upshot is that, for high-quality proxies, concerns about gaming and other strategic behavior by decision-subjects are unlikely to justify either social concerns about disclosure or significant decision-maker resistance to disclosure. The caveat is that there are narrow circumstances in which high-quality proxies are gameable and disclosure could seriously erode their social value. If decision-makers can explain why such circumstances apply to a particular decision, then the social costs of gaming could outweigh the relatively modest expected social value of disclosing to promote follow-on innovation.

*2. Low-Quality Proxies*

Even if information about a low-quality proxy can be used strategically by decision-subjects, it hardly matters because the proxy has so many false negatives that there is no reason to assume the strategic behavior will result in gaming (turning true negatives into false positives) rather than error correction (turning false negatives into true positives). Because a low-quality proxy makes so many errors, it is also unclear whether decision-subjects who strategically adapt their behavior to its requirements can improve their true eligibility for a beneficial decision, so this factor is moot. Overall, gaming is not a persuasive reason to avoid disclosure of a low-quality proxy.

*3. Moderate Quality Proxies*

We argued above that gaming will rarely be a socially substantial rationale for secrecy for either high-quality proxies—which are rarely gameable—or low-quality proxies—for which gaming is holds little meaning. On the flip side, neither socially beneficial error correction nor socially beneficial eligibility improvement is likely to provide a substantial rationale for the disclosure of high- or low-quality proxies.

The situation is different for moderate-quality proxies. Because these proxies make moderate numbers of errors, one can expect that the connection between the proxy and the ideal decision criteria is considerably looser than it is for high-quality proxies. This sort of loose connection is one prerequisite for gaming.

As our prior work also establishes, many aspects of most decision proxies can be disclosed without risk of gaming because they are impossible or costly for decision-subjects to modify strategically.[117] Nonetheless, for some moderate-quality proxies, disclosing the full details of the proxy could lead to socially significant gaming, degrading the quality of the proxy. On the other hand, and for essentially the same reasons, moderate-quality proxies are also more likely than high- or low-quality proxies to provide fertile ground for strategic error correction to change a false negative to a true positive.

If the error profile is symmetric, without significant disparities between groups, then there is no general reason to expect that disclosing details of the proxy is more likely to allow gaming than to facilitate strategic error correction. To be convincing, then, an argument against disclosure based on gaming should explain why gaming is more likely than error correction. The accountability provided by disclosure can also cut down on gaming because the best way to discourage gaming is to improve the fit between the proxy and the ideal decision criteria.

Disclosure might also allow decision-subjects to act strategically to improve their real eligibility for a beneficial decision. Society will ordinarily welcome this behavior, but as discussed earlier, deployers for whom extra beneficial decisions mean a higher workload could strategically resist disclosure to keep down the number of beneficial decisions. One would think, however, that deployers wishing to strategically avoid beneficial decisions would also have procured a proxy skewed to produce extra false negatives. Thus, a symmetric error profile should provide at least some reassurance about this flavor of principal-agent problem.

*4. Moderate-Quality Proxies with Asymmetric or Disparate Error Profiles*

Asymmetric error profiles also provide clues about the likelihood of various sorts of strategic behavior. Suppose, for example, that a proxy produces a small number of false negatives and a large number of false positives. Since there are not many false negatives, strategic error correction is relatively unlikely. The large number of false positives suggests that the connection between the proxy and the ideal decision criteria might be loose enough to allow gaming. On the other hand, the large number of false positives also suggests that the decision-maker is not especially concerned about false positives, suggesting implicitly that gaming (which produces additional false positives) is not a major concern from the decision-makers' perspective.

[117] Cofone & Strandburg, *supra* note 7, at 640–41.

Now, suppose that a proxy's error profile has many false negatives and a small number of false positives. In this case, there could be considerable potential for strategic error correction. The small number of false positives makes a decision-maker's concerns about gaming more plausible because it suggests an effort has been made to avoid the sort of errors that gaming would create.

In the end, much depends on whether the observed asymmetry is aligned with social values. If not, one should worry more about improving the proxy design than about avoiding gaming. If the asymmetry is socially desirable, then its direction also says something about whether gaming is a significant social concern. Inconsistent behavior by the decision-maker is a clue to principal-agent problems. If a designer produces a proxy that generates many false positives, but resists disclosure by asserting gaming concerns, a principal-agent problem should be suspected.

A moderate overall error profile that favors the dominant group suggests that the proxy is better tailored to the characteristics or behavior of that group. Perhaps counterintuitively, the looser connection (and higher error rate) means that strategic behavior may be more feasible for the disfavored group. Even if disclosure leads to some gaming by members of the disfavored group, however, that is reasonable price to pay from a social perspective if disclosure also provides accountability that leads to reduced disparity. If the error profile is both disparate and asymmetric, principal-agent issues in proxy design are even more likely and the social costs of continuing with a biased and sub-optimal decision proxy are even more likely to outweigh social costs of gaming by the disfavored group.

More socially concerning is the possibility that members of the dominant group might use the disclosure to game the system and improve their position even more. The lower rate suggests that strategic behavior is more difficult and costly for the dominant group. Nonetheless, that group might be substantially more well-equipped, financially or otherwise, to game the proxy. In other words, a disclosure aimed at providing accountability and increasing equality could end up advantaging the dominant group (at least in the short run before the proxy's performance for the disfavored group is improved).

Disparate error rates for different social groups suggest principal-agent problems in proxy design that should be investigated. Moreover, generalized assertions about gaming by decision-makers can hide resistance to accountability. Where there is the potential that disclosure to decision-subjects would allow strategic behavior to exacerbate disparities, disclosure regimes should be tailored to take this possibility into account, perhaps by disclosing only non-gameable elements to the public, with more detailed disclosure to an oversight body. There is also the possibility of reducing opportunities for decision-subject strategic behavior by strengthening the connection between the decision proxy and the ideal decision criteria for all social groups.

### *E. General Considerations on Disclosure under Principal-Agent Problems*

The upshot of the discussion so far is that disclosure, at a minimum, of information needed to assess decision quality, gauge the credibility of decision-maker justifications for overall quality and any asymmetries or disparities, should ordinarily be required for socially significant decisions, except perhaps when the decision proxy is of high quality for all social groups. From an innovation perspective, even more detailed disclosure may be warranted depending on the balance between recouping upstream investment and facilitating downstream innovation.

#### *1. High Quality Proxies*

In sum, this analysis suggests that in most cases without commercial implications, society would be best served by full disclosure, particularly when there is the potential for decision-subjects to improve their eligibility for beneficial decisions. The error profile does not suggest a significant need for oversight of proxy design but, since gaming is not a serious concern, there is little reason not to disclose, thus enhancing decision system legitimacy and contributing to the pool of knowledge about decision-making systems. In the limited situations in which decision-makers can explain why gaming is a serious concern, they should also be required to explain why they cannot design a similarly high-quality proxy that can be disclosed without the threat of gaming. In the limited circumstances where gameability cannot be avoided without unacceptably compromising decision quality, disclosure to potential decision-subjects should be avoided. However, it may be socially desirable to set up a disclosure regime involving an oversight body or ombudsman. The use of magistrates to issue warrants in the criminal context is of this ilk.

Where there are commercial trade secrecy interests involved, society and decision-makers will be aligned in preferring some mechanism for allowing developers to recoup their investments. From a social perspective, a tailored disclosure regime, perhaps accompanied by an alternative exclusivity mechanism would be preferable. However, one can expect decision-makers to selfishly resist any significant disclosure, so a mandate will be needed. The design of the disclosure regime for a high-quality proxy need not focus on questions of accountability but should focus on facilitating downstream improvement and dispersion of knowledge. As in the non-commercial case, there may also be situations in which there is serious potential for gaming, and disclosure should be further curtailed.

### *2. Low-Quality Proxies*

In the end, whatever a decision-maker's reason for using a low-quality proxy, disclosure is almost always socially desirable. Disclosure to the public would provide accountability, allowing the public to decide whether the benefits of improving decision quality outweigh the costs of doing so, and will often help uncover principal-agent problems that are impeding improvement. Disclosure to potential competitors will facilitate competition to develop needed improvements in proxy design.[118] And because of the proxy's low quality, one should be unconcerned about the potential for decision-subject gaming.

When decision-makers resist disclosure of a low-quality proxy, one should suspect they have engaged in shirking or other self-serving behavior in proxy design. Assertions that disclosure will lead to socially undesirable gaming or that trade secrecy enhances innovation are not plausible for low-quality proxies. Mandating expansive disclosure of details of a low-quality proxy is usually appropriate.

[118] *See* Katharine Kemp, *Concealed Data Practices and Competition Law: Why Privacy Matters*, 16 Eur. Competition J. 628, 661(2020) ("Concealed data practices hinder privacy-enhancing rivals. . . .").

### 3. *Moderate-Quality Proxies*

Moderate-quality proxies are the most difficult case since there are sometimes reasonable social concerns about gaming or innovation. These are also the cases where principal-agent problems at the proxy design level are most difficult to detect without in-depth auditing. Moderate quality is sometimes explained by designer shirking, but it may also reflect a reasonable balance of cost and benefits for the context. Principal-agent problems in proxy design are much more likely when the error profile is unjustifiably asymmetric or there are disparities between social groups.

Disclosure (or the threat of disclosure) would provide incentives for improvement. Unfortunately, because gaming itself is socially undesirable and dominant groups often have better opportunities for strategic behavior, this problem may not be completely solvable by disclosure. Sometimes, a secret, gameable proxy is the social optimum because it would be too difficult or expensive to devise a sufficiently accurate and unbiased decision proxy that is also ungameable.

A disclosure regime for moderate quality proxies should aim to help policymakers determine whether the moderate quality reflects a socially beneficial tradeoff between various costs and benefits or is a suboptimal consequence of decision-maker strategic behavior. Particularly when the error profile reflects disparities and asymmetries, disclosure should aim to incentivize and facilitate improvement. The disclosure regime should also be tailored, as much as possible, to provide the benefits of disclosure while avoiding serious degradation of the proxy's quality by limiting public disclosure as much as possible to aspects that cannot be gamed. Even so, unlimited trade secrecy is unlikely to be warranted. More often, it will be important to promote innovation through a balanced regime of non-secrecy-based incentive mechanisms and tailored disclosures to potential improvers. Policymakers should keep in mind the fact that decision-makers will often have self-serving reasons to resist disclosure and require justifications for assertions that full secrecy is necessary because of gaming or trade secrecy concerns.

## IV
## Disclosure Design

In this Part, we discuss insights into disclosure design: we explore ways to design disclosure mandates to balance the costs and benefits of disclosure in light of the theoretical contribution above. Disclosure design is important because, as others before us have explained, a difficulty with algorithmic transparency is that "the concepts exist at a fairly abstract, or aspirational level, whereas in order for them to have the tangible effects described above they themselves also need to be more concrete, specific and enforceable."[119] Disclosing details about proxies with sub-optimal error profiles, especially when their outcomes differ between social groups, helps to remedy principal-agent problems and provide the basis for socially aligned innovative improvements, leading to the development of more socially beneficial and less biased proxies. While decision-maker protestations about gaming and trade secrecy are often self-interested and indicative of principal-agent problems, there are situations in which the social costs of gaming and the need to recoup innovative investments should concern society as well. In many situations, it will be possible to design disclosure regimes to illuminate potential bias or other principal-agent problems while keeping enough details of the proxy secret to deter socially undesirable gaming or free riding.

## *A. Structural Components of Disclosure*

### *1. Disclosure is Not Binary*

Discussions of disclosure sometimes proceed as though disclosure were a binary action in which an algorithm is either "disclosed" or "not disclosed."[120] In reality, any disclosure regime involves many choices, including *what* specific information to disclose, *when*, to *whom*, and on what *legal grounds*. Often, many types of information are incorporated into an automated decision-making algorithm, including training data and its sources; input variables or design parameters used to produce outputs; the relative influence of different parameters or components; the overall form of the proxy (often called the "model"); the relative influence of different parameters or components (including, in machine learning, labels and weights afforded to each feature); the system objective (in machine learning, outcome variable); the source code used to compute it; and the relationship between the system objective and the ideal decision criteria.[121]

---

[119] Rebecca Williams et al., *From Transparency to Accountability of Intelligent Systems: Moving Beyond Aspirations*, 4 Data & Pol'y e7, e7–2 (2022).

[120] *See* Christopher S. Yoo, *Beyond Algorithmic Disclosure For AI*, 25 Colum. Sci. & Tech. L. Rev. 314, 316 (2024) (arguing that meaningful transparency requires information on multiple dimensions, not just binary disclosure); *State of the Evidence: Algorithmic Transparency*, Open Gov't P'ship (2023), https://www.opengovpartnership.org/wp-content/uploads/2023/05/State-of-the-Evidence-Algorithmic-Transparency.pdf [https://perma.cc/62DY-6E3R] (noting that some studies operationalize transparency in binary terms, such as bias or no bias, disclosure or no disclosure); Cynthia Rudin et al., *The Age of Secrecy and Unfairness in Recidivism Prediction*, arXiv (2018), https://arxiv.org/abs/1811.00731 [https://perma.cc/FDP4-3Y88] (discussing how the lack of transparency in algorithms can lead to unfair outcomes, highlighting the need for more nuanced approaches beyond binary disclosure).

The question, therefore, is not a binary one of whether to disclose, but rather one of how to design an appropriate disclosure regime.[122] Separate considerations may apply to the social tradeoffs involved in disclosing (or not disclosing) each sort of information to particular recipients using particular legal processes. It is thus useful to have a framework for making choices about disclosure in different decision contexts.[123]

### 2. *Threshold Considerations*

As a first step, two types of disclosure can be required without raising significant gaming or trade secrecy concerns: the distribution of decision outcomes across groups and the error profile.[124] The latter include, for machine learning systems, false positive and false negative rates for the population and significant sub-groups and, for generative or agentic systems, equivalent metrics of directional error and group-level performance disparities.

Disclosing decision proxy error profiles is a threshold step in a well-designed disclosure regime because error profiles allow the public to make an initial assessment of the quality of a proxy and give clues about potential principal-agent problems. At the same time, a regular practice of disclosing error profiles provides incentives for socially aligned innovation.[125] Consequently, error profiles are useful in assessing the validity of decision-maker claims that trade secrecy is necessary

---

[121] Casper et al., *supra* note 74, at 2255–56 (discussing what is entailed in "full access" to the system in a white-box or outside-the-box model); Ryan Calo, *Artificial Intelligence Policy: A Primer and Roadmap*, 51 U.C. DAVIS L. REV. 399, 412–14 (2017) (discussing how automated systems rely on diverse datasets and data integration to function effectively).

[122] Graves & Katyal, *supra* note 6, at 1412; FERGUSON, *supra* note 3, at 139–40.

[123] For example, FOIA's trade secrecy and law enforcement methods exemptions can be quite problematic when used to keep decision-making opaque. *See* Graves & Katyal, *supra* note 6, at 1368–70, 1373, 1376–81, 1415–16.

[124] *See* Cofone & Strandburg, *supra* note 7, at 640–41, 648–51 (discussing the gaming concern).

[125] Kroll et al., *supra* note 3, at 667–69 (arguing that transparency in algorithmic processes, such as error disclosure, incentivizes innovation aligned with public interest); Sandra Wachter et al., *Why a Right to Explanation of Automated Decision-Making Does Not Exist in the General Data Protection Regulation*, 7 INT'L DATA PRIV. L. 76, 91–92 (2017) (highlighting the value of transparency in identifying and addressing

to preserve incentives for innovation. Requiring decision-makers to disclose an error profile, along with information about how it was calculated or estimated, has additional benefits. For example, if a decision-maker using an algorithm fails to disclose an error profile calculated or estimated in a reasonable way, that failure, in and of itself, should raise serious questions about the decision-maker's reliance on that algorithm. Similarly, if a deployer claims not to have any information about the error profile of an algorithm—whether because the designer withheld it or because it was never computed—policymakers might well decide to ban continued use of that algorithm and require that designers supply error profiles as a condition of deployment.

The error profile also provides information that can be used to determine whether and what further disclosure is warranted. Where the error profile suggests the possibility of principal-agent problems in proxy design—especially when there are asymmetries and disparities between social groups—justification should be demanded from the decision-maker. If there is no satisfactory justification, then further disclosure can be required to facilitate accountability and improvement. Given that a proxy with unjustifiable and asymmetric errors will be revised, gaming is a lesser concern. Moreover, if the decision-maker or developer has been behaving strategically, it is reasonable to give competitors an opportunity to design an improved proxy.

In some contexts, a threshold inquiry clarifies that at least some aspects of a decision-making system can be disclosed without significant social tradeoffs.[126] If, for example, the concern is about gaming, consideration of whether to keep information about a decision-making algorithm opaque should begin by asking whether the specific information at issue meets the basic prerequisites for gaming laid out in our earlier analysis: immutable (or effectively immutable) characteristics are ungameable.[127] Also, high-quality proxies are unlikely to be gameable unless they have incidental (for example, temporal) connections to decision characteristics of interest that make them accurate for some situations but not others.[128] However, algorithmic errors to improve societal outcomes). *See also* Qiang Chen et al., *Financial Disclosure and Innovation*, 58 J. Acct. Res. 865, 865–900 (2020).

[126] *See* Cofone & Strandburg, *supra* note 7, at 650.

[127] *See supra* Section I.A.

[128] *See supra* Section I.A.

if the proxy is of very low quality, the benefits of disclosure will nearly always outweigh concerns about gaming or trade secrecy.

When a decision-maker or developer asserts trade secrecy concerns, it is important to consider whether all of the assertedly protected information doctrinally qualifies as trade secret subject matter.[129] Even when all the information arguably qualifies as a trade secret, there are various circumstances in which concerns about discouraging innovation are minimal either because there is no competitive market (as is sometimes the case with government development or procurement of decision-making systems) or because features of the particular market disadvantage potential free riders.[130] Because "misappropriation" is required for a trade secrecy violation, government-mandated disclosure never constitutes an actionable trade secrecy violation. Instead, the question at hand is whether mandating disclosure is likely to discourage socially beneficial innovation.[131]

In sum, as a threshold matter, considerable information about a decision-making system can often be disclosed to the public without the need to delve deeply into tradeoffs between disclosure's social benefits and the potential social costs of gaming or competitor free riding.

### *B. Types of Information to Disclose*

#### *1. Low-Risk Disclosures*

First and foremost, information about error profiles and the distribution of decision outcomes across demographic groups is useful in identifying principal-agent problems and poses no legitimate gaming or trade secrecy concerns. Thus, in socially important contexts, designers should be required to calculate and provide an error profile, and deployers should be required to obtain and disclose both, along with, for machine learning algorithms, statistical information about the representativeness of the training and test data to assess the error profile. Society's losses from gaming and trade secrecy breaches, and society's benefits from better oversight of decision-makers, both must be evaluated in light of the specific disclosures that can be made.[132] As we noted, many types of information can be useful in holding decision-makers accountable and uncovering principal-agent problems, but information on error profiles is particularly informative.[133]

---

[129] John Villasenor, *Artificial Intelligence, Trade Secrets, and the Challenge of Transparency*, 25 N.C. J.L. & TECH. 495, 526 (2024); *Intellectual Property Toolkit–Trade Secrets*, U.S. Patent & Trademark Office (2023), https://www.uspto.gov/sites/default/files/documents/tradesecretsiptoolkit.pdf. [https://perma.cc/X86B-FZ6Q]

[130] Wexler, *supra* note 6, at 1419–20, 1422–23; Yafit Lev-Aretz & Katherine J. Strandburg, *Privacy Regulation and Innovation Policy*, 22 YALE J.L. & TECH. 256, 301–02 (2020); Siems et al., *supra* note 40, at 812–13.

[131] Coglianese, *supra* note 37, at 236, 240–43; Citron & Pasquale, *supra* note 4, at 26.

One objection is that computing these error profiles will be burdensome for decision-makers. Error rates, in particular, often can only be computed using available data (often the training data or test data) and evaluation methods, so they may not always be accurate with respect to the metrics that society cares about.[134] Nonetheless, AI designers are well-positioned to compute these profiles, as they curate training data, collect data about model performance, fine-tune their models, and are overall best-positioned to estimate and correct error asymmetries. For that reason, decision-makers and developers should be required to use best practice methodologies to compute approximate error profiles that can be used as first steps in assessing whether principal-agent problems are likely to arise and to disclose both the estimated error rates and the methods used to derive them.

A related objection is that error rates might simply be unknown, either because the decision-maker failed to compute them or because ground-truth outcomes are unavailable. Three considerations apply to this situation. First, the unavailability of error profile information is itself diagnostic: a decision-maker who cannot produce an error profile for a consequential algorithm has failed the minimum standard the framework requires. Second, the demographic composition of affected populations and the distribution of outcomes across groups are observable without requiring ground-truth error computation and support rebuttable presumptions of principal-agent problems when disparities are evident. Third, the disclosure mandate is not purely diagnostic: it is also designed to generate information. By mandating disclosure of population distributions and error profiles as conditions of deployment, the framework creates, rather than presupposes, the informational infrastructure needed for accountability.

---

[132] *See* Villasenor, *supra* note 129, at 511, 516 (arguing against a broad reading of AI trade secrets and tailoring such claims to specific descriptions of misappropriated algorithms); PASQUALE, *supra* note 4, at 142.

[133] *See supra* Part III.

[134] Cofone & Khern-am-nuai, *supra* note 84, at 1459.

Society often benefits from mandating tailored disclosures that go beyond error profiles and beyond other information that cannot raise gaming or trade secrecy concerns. It is often possible to mitigate concerns about gaming or suppressing innovation by mandating disclosures about the proxy at a coarsened level of generality or disclosure of information about validation and testing protocols and their results.[135] For example, if Amazon had developed an infallible machine learning model to identify the best engineers in the industry, it may not want Google or Meta to have access to the model's inner workings, but disclosing its data, sources, and labels would hardly give rise to such a concern.

### 2. *Tailored Disclosures*

Additional tailored disclosures are feasible and useful in most decision-making contexts. Because disclosures are not all-or-nothing, there are many things a decision-maker can disclose while not disclosing to others. Chiefly, decision-makers can choose to disclose the model, system objective, design parameters (such as features and weights in machine learning systems), code, training data, data sources, demographics, dataset size, validation protocols, and the results of validation protocols.[136] The specific types of disclosures should depend on the application and the stakeholders involved so that the disclosures are both meaningful in the given context. Here, we propose four elements that should be adjusted in tailored disclosures: the system's general characteristics, the data, the code, and the model.

The first, most cautionary, tailored approach is the disclosure of high-level descriptions. Instead of revealing detailed code or proprietary information, decision-makers can provide high-level descriptions of their algorithms, focusing on the overall methodology, principles, and concepts behind the algorithms without disclosing specific implementation details.[137] These high-level descriptions enable stakeholders to evaluate whether design choices align with stated policy objectives without exposing sensitive implementation details: they can provide socially valuable information about the algorithms' functioning without revealing trade secrets or allowing for gaming.

---

[135] PASQUALE, *supra* note 4, at 142.

[136] Casper et al., *supra* note 74, at 2255–56 (disclosing "white-box" and "outside-the-box" access regimes); Werder et al., *supra* note 74, at 22:2–3, 22:14–15.

[137] Coglianese & Lehr, *supra* note 3, at 1208–12; Kroll et al., *supra* note 3, at 662–74. *See also* Selbst & Barocas, *supra* note 4, at 1113–15; Lipton, *supra* note 9, at 15–20.

High-level descriptions, for example, might be useful for knowing whether fairness constraints are generally being applied in the decision process. They also allow stakeholders (e.g., regulators, consumers, and organizations) to understand the general framework and decision-making principles guiding the algorithm, which may be appropriate for low-risk applications. Their helpfulness, however, faces significant limitations. While general descriptions provide insight into underlying principles, detailed data inputs, design parameter choices (e.g., feature selection), and model-specific behaviors are necessary to evaluate bias and discrimination. For example, if a hiring algorithm claims to consider the principle of "fairness," a high-level description alone will not reveal whether the training data disproportionately favors one demographic group.[138] Similarly, auditors and third-party evaluators require access to the code, data, and implementation details to conduct reproducibility checks and audits.[139]

The second tailored approach is the disclosure of certain characteristics of the input data in contexts where releasing all data is unfeasible. One such disclosure is information about the sources of the data used to train and test the models, including statistics about its representativeness (called data provenance or data lineage).[140] This sort of information can be used for data quality assessments. For example, if an algorithm is used to predict medical diagnoses, disclosing data sources (e.g., hospital records, lab tests) and their representativeness (e.g., age, gender, and geographic diversity) allows users to assess whether the data accurately reflects the population it's intended to serve.[141] Data sources and representativeness also affect

[138] Further, the precise definition of "fairness" itself has been fiercely contested. *See, e.g.*, Alexandra Chouldechova & Aaron Roth, *The Frontiers of Fairness in Machine Learning*, ARXIV (Oct. 20, 2018), https://arxiv.org/abs/1810.08810 [https://perma.cc/3LGN-32JL]; Angwin et al., *supra* note 32.

[139] Casper et al., *supra* note 74, at 2257–58; Werder et al., *supra* note 74, at 22:14.

[140] Werder et al., *supra* note 74, at 22:7–8; Julia Stoyanovich & Bill Howe, *Follow the Data! Algorithmic Transparency Starts with Data Transparency*, ETHICAL MACH. (Nov. 27, 2018), https://web.archive.org/web/20230926064209/https://ai.shorensteincenter.org/ideas/2018/11/26/follow-the-data-algorithmic-transparency-starts-with-data-transparency.

[141] Anmol Arora et al., *The Value of Standards for Health Datasets in Artificial Intelligence-Based Applications*, 23 NAT. MED. 2329, 2935 (2023) ("A key theme found through the systematic review was a need for transparency in how datasets are prepared, including who is included or excluded from the dataset, how missing data are handled and how data are labeled. Greater transparency in these areas allows better

how well a model generalizes to unseen data. By making this information available, third parties can identify issues such as poor generalization or systematic bias caused by unrepresentative or narrow data sources. For example, if an AI model trained on a limited dataset performs poorly when applied to new data, transparency about the data's scope and limitations can help stakeholders see how to improve the model by acquiring additional, more representative data. In an AI system for child welfare case management, disclosing that data comes from multiple sources (e.g., state databases, court records, and social worker notes) can help users assess whether certain types of cases or demographics are overrepresented.[142]

Similarly, decision-makers can disclose aggregated data or statistics generated by the algorithm without revealing specific algorithmic details.[143] Doing so can provide transparency about the overall performance and outcomes of the algorithm without compromising proprietary information or facilitating gaming.[144] The most important of these are the distribution of decision outcomes across groups and the error profile, since, as discussed above, they serve as a diagnostic tool for principal-agent problems.[145] For example, in the context where an AI system is used for evidence in criminal trials, the company can disclose aggregated statistics, like the proportion of false positives and false negatives for different demographic groups, without revealing how exactly the algorithm reaches its determinations.[146] By receiving aggregate metrics on the performance of an algorithm across different understanding of the context and limitations of a dataset, which in turn provides a guide to the potential limitations of any inferences or innovations derived from that dataset"); Natalia Norori et al., *Addressing Bias in Big Data and AI for Health Care: A Call for Open Science*, 2 PATTERNS 100347 (2021).

[142] *See* Devansh Saxena & Shion Guha, *Algorithmic Harms in Child Welfare: Uncertainties in Practice, Organization, and Street-Level Decision-Making*, ARXIV (Aug. 9, 2023), https://arxiv.org/abs/2308.05224 [https://perma.cc/8XBM-AR8Z] (raising "serious data provenance concerns about data collected about children through the [Child and Adolescent Needs and Strengths] algorithm, since the data is so heavily manipulated by both caseworkers and foster parents").

[143] ALFRED RADAUER ET AL., STUDY ON THE LEGAL PROTECTION OF TRADE SECRETS IN THE CONTEXT OF THE DATA ECONOMY 1, 39–40, 78–80 (2022); Burrell, *supra* note 34, at 10. Some such information may be found in model cards. *See, e.g.*, *GPT-4o System Card*, OPENAI (Aug. 8, 2024), https://openai.com/index/gpt-4o-system-card/?ref=maginative.com [https://perma.cc/W7B3-WCDC].

[144] Citron & Pasquale, *supra* note 4, at 26.

[145] *See supra* Part III.

[146] Rebecca Wexler, *It's Time to End the Trade Secret Evidentiary Privilege Among Forensic Algorithm Vendors*, BROOKINGS (July 13, 2021), https://www.brookings.edu/articles/its-time-to-end-the-trade-secret-evidentiary-privilege-among-forensic-algorithm-vendors/ [https://perma.cc/J26N-XDPY]; Angwin et al., *supra* note 32.

demographic groups (e.g., race, gender, socioeconomic status), stakeholders can assess whether the algorithm is disproportionately affecting any specific group.

The third tailored approach is the disclosure of redacted code. Decision-makers can redact or remove sensitive portions of algorithmic code or other proprietary information while still providing some understanding of the algorithm's functioning.[147] This can involve removing specific lines of code, functions, or proprietary data while retaining the general structure or logic of the algorithm. Disclosing redacted code allows the public, auditors, and other stakeholders to understand the basic structure and logic of the algorithm without exposing proprietary methods or sensitive business information. Redacted code can reveal the flow of data, high-level methodologies, and key decision points in the algorithm while protecting trade secrets, thus improving transparency. This sort of redaction is not always possible or meaningful but is available in some contexts. For example, in a machine learning hiring system, redacted code could show how the algorithm processes input data (e.g., resumes, qualifications) without disclosing the specific weightings or proprietary elements that lead to the final prediction. Similarly, redacted code from a credit scoring system can show how the algorithm processes inputs (e.g., payment history, credit inquiries) and calculates scores, allowing auditors to spot potential areas where certain groups might be unfairly penalized.

Finally, aspects of the model can be disclosed in a tailored way without risking gaming or trade secrecy infringements. An important type of disclosure is explaining all intended uses of the algorithm, detailing when and where the model should and should not be applied.[148] This type of disclosure allows stakeholders to verify that decision-makers use the algorithm in appropriate contexts and makes them aware of limitations that might make the model less effective or reliable in certain scenarios.[149] This type of disclosure can be helpful in predictive policing, loan approval, and hiring systems.[150] For example, a machine learning recruitment

[147] Kroll et al., *supra* note 3, at 659.

[148] *See* Johann Laux et al., *Three Pathways for Standardization and Ethical Disclosure by Default Under the European Union Artificial Intelligence Act*, 53 COMP. L. & SEC. REV. 105957 (2024).

[149] *See* Rebecca Williams, *Social Scoring and the Law's Response*, Part II (draft 2025, on file with author). *Cf.* Alexander J. Wulf & Ognyan Seizov, *"Please Understand We Cannot Provide Further Information": Evaluating Content and Transparency of GDPR-Mandated AI Disclosures*, 39 AI & SOC'Y 235, 246–47, 251 (2024) (finding such disclosures are often vague, incomplete, or not actionable).

[150] *See, e.g.*, FERGUSON, *supra* note 3, at 136–40.

AI might disclose that its predictions are optimized for candidates with traditional educational backgrounds but may not perform as well for candidates with non-traditional careers. Another is disclosing how the model will be monitored after its deployment to ensure continued performance, together with the process by which it will be updated or maintained. AI models can degrade over time as the contexts they were designed for evolve and the data they use becomes outdated.[151] Understanding its monitoring can be helpful for stakeholders in applications such as lending, where decision-making needs to adapt to changes in patterns or behavior. A financial risk assessment AI, for example, might disclose that it will monitor decision quality over time and update or recalibrate the model periodically based on new data, or introduce a regular third-party audit process to track model performance.

### *C. Forms of Disclosure*

#### *1. Audiences: Disclosure to Whom*

When "the public" is the principal—as is generally the case for decision-making systems—it is natural to begin by thinking of disclosure to the public as the most effective way to obtain accountability. Disclosure to the public has many advantages and should be considered the gold standard for disclosure design. However, given that decision-subjects (potential gamers) and competitors (potential free riders) are also members of the public, it may be necessary to consider alternative audiences for some sorts of disclosures—particularly at the most detailed level.

Another reason for considering alternative audiences is that some types of disclosures to the general public, while harmless, are unhelpful because they cannot be interpreted without some degree of technical expertise.[152] These can be either technical expertise about the AI system's design and operation—whether machine learning, generative AI, or agentic systems—or expertise about the underlying decision-making context.[153] For example, the public might not be able to do much with redacted code. In those situations, alternative audiences should be considered *in addition to* the principal audience of the general public.

---

[151] *See, e.g.*, Daniel Vela et al., *Temporal Quality Degradation in AI Models*, 12 SCI. REP. 11654 (2022). *See also* Ilia Shumailov et al., *AI Models Collapse When Trained on Recursively Generated Data*, 631 NATURE 755, 757–58 (2024).

[152] *See* Wulf & Seizov, *supra* note 149, at 243–45; Jesse Eisinger, *The Trouble with Disclosure: It Doesn't Work*, PROPUBLICA (Feb. 11, 2015), https://www.propublica.org/article/the-trouble-with-disclosure-it-doesnt-work [https://perma.cc/MGM8-ECLH].

Building on pre-AI disclosure regimes, there are a variety of alternatives or complements to disclosing information to the general public. Three stand out.

One audience alternative is enabling independent audits.[154] Organizations can engage third-party auditors to conduct independent audits of their algorithms.[155] These auditors may review the algorithm's performance, fairness, and so on, while respecting the organization's trade secrets and not disclosing information that enables gaming by decision-subjects.[156] The audit reports can then be used to provide transparency to stakeholders without disclosing proprietary information or facilitating gaming.[157]

A second audience alternative is conducting an expert agency review. Many are unpersuaded by proposals for a specific agency for algorithms or AI (a sort of "FDA for algorithms)"[158] for reasons of practicality and risk of capture. However, the idea of disclosing some types of information to an expert government agency is more familiar and less controversial. Disclosure to expert agencies is reflected in, for example, proposals for reporting "algorithmic impact assessments" to oversight bodies.[159]

---

[153] *See* FERGUSON, *supra* note 3, at 194.

[154] *See, e.g.*, C-203/22, CK v. Magistrat de Stadt Wien, *supra* note 57, at ¶ 76 (allowing a trusted third party access to a supervisory authority).

[155] *See generally* Gregory Falco et al., *Governing AI Safety Through Independent Audits*, 3 NAT. MACH. INTEL. 566, 567 (2021); Inioluwa D. Raji et al., *Outsider Oversight: Designing a Third Party Audit Ecosystem for AI Governance*, 5TH ANNUAL ACM/AAAI AI ETHICS & SOC'Y CONF. 557, 562–63 (2022).

[156] *See* Bruno Lepri et al., *Fair, Transparent, and Accountable Algorithmic Decision-Making Processes*, 31 PHIL. & TECH. 611, 620–21 (2018); Kroll et al., *supra* note 3, at 661; FERGUSON, *supra* note 3, at 197–98.

[157] *See* Michael Veale et al., *Fairness and Accountability Design Needs for Algorithmic Support in High-Stakes Public Sector Decision-Making*, ARXIV, at 5 (Feb. 3, 2018), https://arxiv.org/abs/1802.01029 [https://perma.cc/G5TL-JJEJ] (discussing internal audits).

[158] *See* Tutt, *supra* note 78, at 104–11. *See also The World Needs an International Agency for Artificial Intelligence, Say Two AI Experts*, ECONOMIST (Apr. 18, 2023), https://www.economist.com/by-invitation/2023/04/18/the-world-needs-an-international-agency-for-artificial-intelligence-say-two-ai-experts [https://perma.cc/D7XT-2FN3].

[159] *See, e.g.*, Margot E. Kaminski & Gianclaudio Malgieri, *Multi-Layered Explanations from Algorithmic Impact Assessments in the GDPR*, 3 PROC. OF 2020 CONF. ON FAIRNESS, ACCOUNTABILITY, & TRANSPARENCY 68, 72 (2020). *See also* Paul B. de Laat, *Algorithmic Decision-Making Based on Machine Learning from Big Data: Can Transparency Restore Accountability?*, 31 PHIL. & TECH 525 (2018).

A third alternative is disclosure to specific parties or the court itself during a trial, for example under a protective order. In litigation, disclosure of trade secret information is often made under a protective order that restricts the audience to an opposing party's legal representatives or experts, or more broadly to those involved in the trial, excluding journalists and the public from the audience.[160] While often used to protect trade secrets, protective orders can also be used to protect against gaming. In-camera review, where a judge examines the information, as is often done for issues of national security serves an equivalent, and even more restrictive function.[161] Analysis of information disclosed under a protective order or during an in-camera hearing can then determine the extent of a broader disclosure. Outside of litigation, other sorts of trusted representatives, such as unions or public interest organizations, might be enlisted to perform a similar protective function.

Oversight-only disclosure is appropriate when public release would meaningfully increase gaming or undermine innovation, yet a principal-agent problem requires evaluation of algorithmic design. Generally, these alternative audiences function as intermediaries who improve the social tradeoffs of disclosure by representing "the public" from an accountability perspective while keeping "decision-subjects" and "competitors" in the dark as to gaming and free riding, respectively. However, introducing such intermediaries also introduces new points where principal-agent problems can arise. Therein lies the rub. There are no perfect solutions to this dilemma, but choosing intermediaries appropriately can at least ensure that, due to principal-agent problems, intermediaries do not exacerbate the problems they are intended to solve. Thus, for example, the expert agency approach makes the most sense when seeking to hold private actors to account; independent auditors could be used for either government or private actors, but attention should be paid to how their work is funded; and group representatives should perhaps have fiduciary duties of loyalty to those they represent. In any event, these problems, while serious, are, for the most part, well-studied and not especially novel.

---

[160] Wexler, *supra* note 6, at 1423 ("Compelled disclosure subject to a protective order is thus hardly guaranteed to produce the adverse results for innovation some developers have claimed.").

[161] Uniform Trade Secrets Act, § 5 ("[A] court shall preserve the secrecy of an alleged trade secret by reasonable means, which may include granting protective orders in connection with discovery proceedings, holding in-camera hearings, sealing the records of the action, and ordering any person involved in the litigation not to disclose an alleged trade secret").

There is, however, one aspect of automated decision-making that raises novel concerns. The evaluation of machine-learning-based decision-making systems may require expertise not only about the substantive arena in which the proxies are used, but also about the data science underlying their development. Often, this division of expertise introduces an additional layer of potential principal-agent problems between decision-makers and proxy developers.[162] It may also complicate the design of effective "alternative audience" regimes, given that the agencies, auditors, or group representatives may need dual expertise, as well as access to information that may not be available to the decision-makers themselves. This issue, while important, also affects the efficacy of public disclosure. After all, the public's ability to coordinate data science and subject matter expertise is likely to depend on the decision context—and especially on the resources available to those with "skin in the game," presumably decision-subjects. Thus, even if public disclosure is socially desirable, it will still be important to think about the role of intermediaries in putting that disclosure to use for demanding accountability.

### 2. *Mechanisms: Disclosing under What Legal Grounds*

A mandated disclosure regime can also employ a variety of legal processes to regulate what gets disclosed to whom under what circumstances. We discuss three notable examples.

The first is using audience-limiting mechanisms such as non-disclosure agreements (NDAs) and the abovementioned protective orders.[163] When disclosing to specific audiences besides the general public, decision-makers can require stakeholders, such as researchers, partners, and sometimes even business customers, to sign NDAs that bind them to confidentiality regarding proprietary information.[164] In the case of disclosing to a court, this can be done under a protective order. Doing so provides a mechanism for protecting trade secrets and preventing gaming while allowing for controlled access to information about the decision-making process, thereby bringing in some of the benefits of disclosure.[165] NDAs can allow for trusted third party access in expert audits.[166]

[162] *See* Cofone & Strandburg, *supra* note 7, at 638 (discussing these issues at the interface between decision-makers and tool developers); Sabine Gless, *AI in the Courtroom: A Comparative Analysis of Machine Evidence in Criminal Trials*, 51 Geo. J. Int'l L. 195, 211 (2020) ("Even experts called upon to explain machine evidence in court encounter limitations in their ability to comprehensibly explain how an AI-driven device evaluates a human user's conduct or demonstrate a clear chain of causality.").

[163] *See, e.g.*, Hrdy & Seaman, *Beyond Trade Secrecy: Confidentiality Agreements that Act Like Noncompetes*, 133 Yale L.J. 669, 684 (2024).

[164] Citron & Pasquale, *supra* note 4, at 28 ("Even if scorers successfully press to maintain the confidentiality of their proprietary code and algorithms vis-à-vis the public at large, it is still possible for independent third

The second is filing for patent and copyright protection for some aspects of the AI. Organizations can file for patents for specific elements of their decision-making systems, such as the model, to disclose it while protecting themselves from risk from competitors. Patents provide wider protection for innovations than trade secrets do by granting exclusive rights enforceable against all parties, including those who independently develop the same invention.[167] They therefore allow organizations to disclose elements of their automated decision-making processes in patent applications while still maintaining trade secrecy for other aspects.[168] Intellectual property tools such as patents, often used to safeguard industrial secrets, can also prevent gaming if an organization patents (and thus discloses) the less gameable aspects of an AI. Software is also copyrightable to some extent and, while copyright's protection of "expression" will not stop competitors from independently implementing an algorithm, it can be used to target blatant copying of things such as source code and human-computer interfaces. Disclosure requirements, which would presumably apply also to competitors, would facilitate the detection of patent and copyright infringement.

The third is implementing knowledge governance arrangements such as phased disclosure. With phased disclosure, progressively more granular details are made available if and when initial disclosures are shown to be questionable parties to review it. One possibility is that in any individual adjudication, the technical aspects of the system could be covered by a protected order requiring their confidentiality.").

[165] *See* Wexler, *supra* note 6, at 1420–21. *But see* Katyal, *supra* note 34, at 126–29 (explaining that whistleblowing is an effective way to increase algorithmic transparency).

[166] *See, e.g.*, C-203/22, CK v. Magistrat de Stadt Wien, *supra* note 57, at ¶ 76 (allowing a trusted third party access to a supervisory authority).

[167] Lionel Bently & Tanya Aplin, *Patents and Trade Secrets*, *in* Overlapping Intellectual Property Rights 65, 66, 72–73 (Neil Wilkof ed., 2d ed. 2023); Katherine Linton, *The Importance of Trade Secrets: New Directions in International Trade Policy Making and Empirical Research*, 36 J. Int'l Trade & Econ. 1, 4 (2016).

[168] Jeanne C. Fromer, *Dynamic Patent Disclosure*, 69 Vand. L. Rev. 1715, 1716–18 (2016) (highlighting how companies can disclose certain elements while keeping others confidential); Lisa Ouellette, *Do Patents Disclose Useful Information?*, 25 Harvard J.L. & Tech. 1, 558, 575–77, 585–88 (2012) (noting that companies may choose to disclose only certain aspects of their inventions).

or problematic.[169] This approach could require some prima facie showing by those challenging the proxy, but it does not need to. It would be equally possible (and often preferable) to begin by requiring a set of mandatory disclosures, at least regarding some non-gameable and non-free-rideable information such as data sources, data representativeness, and system objectives. Further levels of disclosure could then depend on the extent to which earlier disclosures suggested problems. Related to phased disclosures are knowledge governance arrangements known as "walled gardens."[170] Usage restrictions (and even NDAs) can be combined with technical measures to create governance regimes that limit not only the audiences for disclosure, but also the uses to which the disclosed information can be put. For example, access to information about a decision-making system might be limited to a specific set of qualified researchers, who could be limited to certain types of testing.

## Conclusion

Algorithmic decision-making systems shape critical aspects of people's lives, from hiring to credit allocation. While these systems promise efficiency, deliberate opacity conceals biases and socially harmful practices. This article shows that principal-agent problems—where decision-makers prioritize their private interests over the public good—are central to understanding algorithmic opacity. It advances a framework for determining when algorithmic transparency should be mandated, even when there are concerns about gaming the system or protecting trade secrets.

Genuine tradeoffs exist. Fears about gaming and trade secrecy are often overstated or strategically deployed to resist accountability. But the tradeoffs should nevertheless be taken seriously. Algorithm performance is determined not only by accuracy but also by how tradeoffs between different types of errors are affected by principal-agent problems. Even when gaming or free-riding concerns are genuine, an appropriately structured disclosure regime is less socially costly than opacity.

---

[169] *See generally* Citron & Pasquale, *supra* note 4 (discussing free disclosure).

[170] Megan Kirkwood, *Regulating the Walled Garden: The Challenge of Taking on the Gatekeepers*, Tech Pol'y Press (Feb. 6, 2024), https://www.techpolicy.press/regulating-the-walled-garden-the-challenge-of-taking-on-the-gatekeepers/ [https://perma.cc/DWF2-CCK5]. *See also* Jeanette Herrle, *Tearing Down a Tech Giant's Walled Garden*, Ctr. For Int'l Gov't Innovation (Oct. 18, 2019), https://www.cigionline.org/articles/tearing-down-tech-giants-walled-garden/ [https://perma.cc/4RKS-PVK7].

Four mechanisms drive the divergence between private disclosure incentives and socially optimal ones: (1) design-stage underinvestment or bias, (2) strategic resistance to disclosure, (3) failure to internalize the benefits of decision-subject improvements, and (4) overvaluation of exclusivity. Together, they illustrate both tiers of principal–agent problems: those arising in algorithmic design and those arising in disclosure decisions.

Error profiles—patterns in the types of mistakes algorithms make—and the distribution of decision outcomes across groups are diagnostic tools for detecting principal-agent problems. These profiles can reveal whether decision-makers are designing proxies that align with societal values or skewing them to serve private interests. For example, error asymmetries that disproportionately harm marginalized groups or disparities in decision quality between social groups often signal the need for greater transparency and oversight. Where error rates are unavailable, the distribution of outcomes across groups serve a subsidiary function and can support a rebuttable presumption of principal-agent problems when disparities are marked. By disclosing these, society can more easily hold decision-makers accountable, improve decision quality, and facilitate fair treatment across populations.

The article also critiques a pervasive justification for AI secrecy: appeals to technical inscrutability. While some machine learning models are complex, meaningful disclosure about error patterns, training data, and key system design parameters is feasible and vital for public accountability. Technical opacity should not serve as an excuse to evade scrutiny, particularly when disclosure mechanisms can be tailored to balance transparency with legitimate concerns about gaming or innovation.

Ultimately, this article presents a novel approach to algorithmic transparency that favors a context-sensitive disclosure regime that balances transparency's social benefits with its potential costs. By translating principal–agent analysis into guidance for regulators, courts, and policymakers, it offers a framework for evaluating AI secrecy claims, structuring disclosure obligations, and allocating oversight. In doing so, we highlight the need to foster accountability so that algorithmic systems serve societal interests rather than private ones.